\documentclass[lettersize,journal]{IEEEtran}

\usepackage[%
    font={normalsize,sf},
    labelfont=sf,
    format=hang,    
    format=plain,
    margin=0pt,
    width=0.8\textwidth,
]{caption}
\usepackage[list=true]{subcaption}
\usepackage{url}
\usepackage{verbatim}

\usepackage[utf8]{inputenc}
\usepackage[T1]{fontenc}
\usepackage[scaled=0.92]{helvet}  
\usepackage[noadjust]{cite}

\usepackage{graphicx}
\usepackage{array}
\newcolumntype{M}[1]{>{\centering\arraybackslash}m{#1}}
\usepackage{tabularx,booktabs}
\newcolumntype{C}{>{\centering\arraybackslash}X} 
\usepackage{dblfloatfix}
\usepackage{amsmath,amssymb}
\usepackage{newtxtext, newtxmath}
\usepackage{dsfont}
\usepackage{hhline,booktabs}
\usepackage{tablefootnote}
\usepackage{enumitem}
\usepackage[T1]{fontenc}
\usepackage{adjustbox}
\usepackage{booktabs, multirow}
\usepackage{amsmath}

\usepackage[scaled=0.9]{DejaVuSansMono}
\usepackage{listings}
\usepackage{amsfonts}

\usepackage{amsthm,thmtools,xcolor}
\declaretheoremstyle[
  headfont=\color{red}\normalfont\bfseries,
  bodyfont=\color{red}\normalfont\itshape,
]{colored}

\usepackage{bm}
\usepackage{url}
\usepackage{mathtools}
\usepackage{balance}
\allowdisplaybreaks

\usepackage{algorithm}
\usepackage{algpseudocode}
\usepackage{etoolbox}
\usepackage[mathscr]{euscript}
\usepackage{calc}
\usepackage{pgfplots}
\usepackage{tikz}
\usepackage{setspace}
\usepackage{pgfplotstable}

\usepackage[font=small]{caption}
\DeclareMathAlphabet{\pazocal}{OMS}{zplm}{m}{n}
\DeclareSymbolFont{missing}{OML}{cmr}{m}{n}
\DeclareMathSymbol{\ell}{\mathord}{missing}{'140}

\usetikzlibrary{calc}
\usetikzlibrary{patterns}
\usetikzlibrary{spy,backgrounds}
\pgfplotsset{grid style={dotted,gray}}

\newcommand{\minimize}{%
  \mathopen{}\operatorname*{minimize}%
}

\newcommand{\subjto}{\textup{subject to}}

\newcounter{problem}
\newcounter{save@equation}
\newcounter{save@problem}

\newtheorem{lemma}{Lemma}

\newlength{\depthofsumsign}
\newlength{\totalheightofsumsign}
\newlength{\heightanddepthofargument}

\makeatletter
\tikzset{reset label anchor/.code={%
    \let\tikz@auto@anchor=\pgfutil@empty
    \def\tikz@anchor{#1}
  },
  reset label anchor/.default=center
}

\let\save@mathaccent\mathaccent
\newcommand*\if@single[3]{%
  \setbox0\hbox{${\mathaccent"0362{#1}}^H$}%
  \setbox2\hbox{${\mathaccent"0362{\kern0pt#1}}^H$}%
  \ifdim\ht0=\ht2 #3\else #2\fi
}
\newcommand*\rel@kern[1]{\kern#1\dimexpr\macc@kerna}
\newcommand*\widebar[1]{\@ifnextchar^{{\wide@bar{#1}{0}}}{\wide@bar{#1}{1}}}
\newcommand*\wide@bar[2]{\if@single{#1}{\wide@bar@{#1}{#2}{1}}{\wide@bar@{#1}{#2}{2}}}
\newcommand*\wide@bar@[3]{%
  \begingroup
  \def\mathaccent##1##2{%
    \let\mathaccent\save@mathaccent
    \if#32 \let\macc@nucleus\first@char \fi
    \setbox\z@\hbox{$\macc@style{\macc@nucleus}_{}$}%
    \setbox\tw@\hbox{$\macc@style{\macc@nucleus}{}_{}$}%
    \dimen@\wd\tw@
    \advance\dimen@-\wd\z@
    \divide\dimen@ 3
    \@tempdima\wd\tw@
    \advance\@tempdima-\scriptspace
    \divide\@tempdima 10
    \advance\dimen@-\@tempdima
    \ifdim\dimen@>\z@ \dimen@0pt\fi
    \rel@kern{0.6}\kern-\dimen@
    \if#31
    \overline{\rel@kern{-0.6}\kern\dimen@\macc@nucleus\rel@kern{0.4}\kern\dimen@}%
    \advance\dimen@0.4\dimexpr\macc@kerna
    \let\final@kern#2%
    \ifdim\dimen@<\z@ \let\final@kern1\fi
    \if\final@kern1 \kern-\dimen@\fi
    \else
    \overline{\rel@kern{-0.6}\kern\dimen@#1}%
    \fi
  }%
  \macc@depth\@ne
  \let\math@bgroup\@empty \let\math@egroup\macc@set@skewchar
  \mathsurround\z@ \frozen@everymath{\mathgroup\macc@group\relax}%
  \macc@set@skewchar\relax
  \let\mathaccentV\macc@nested@a
  \if#31
  \macc@nested@a\relax111{#1}%
  \else
  \def\gobble@till@marker##1\endmarker{}%
  \futurelet\first@char\gobble@till@marker#1\endmarker
  \ifcat\noexpand\first@char A\else
  \def\first@char{}%
  \fi
  \macc@nested@a\relax111{\first@char}%
  \fi
  \endgroup
}

\def\endthebibliography{%
  \def\@noitemerr{\@latex@warning{Empty `thebibliography' environment}}%
  \endlist
}

\algnewcommand{\LineComment}[1]{\Statex \hskip\ALG@thistlm
  \(//\) #1}
\def\BState{\State\hskip-\ALG@thistlm}

\algdef{SE}[DOWHILE]{Do}{EndDoWhile}{\algorithmicdo}[1]{\algorithmicwhile\ #1}%

   \tikzset{nomorepostaction/.code=\let\tikz@postactions\pgfutil@empty}

   \long\def\ifnodedefined#1#2#3{%
   \@ifundefined{pgf@sh@ns@#1}{#3}{#2}%
 }
\usetikzlibrary{spy}
\usetikzlibrary{decorations.pathmorphing,arrows.meta,positioning}
\tikzstyle{printersafe}=[decoration={amplitude=0pt}]

\def\ps@IEEEtitlepagestyle{
  \def\@oddfoot{\mycopyrightnotice}
  \def\@evenfoot{}
}
\def\mycopyrightnotice{
  {\footnotesize
  \begin{minipage}{\textwidth}
  \centering
Copyright (c) 2021 IEEE. Personal use of this material is permitted. However, permission to use this material for any other purposes must be obtained from the IEEE by sending a request to pubs-permissions@ieee.org.
  \end{minipage}
  }
}

\makeatother

\usetikzlibrary{automata}
\usetikzlibrary{calc,fit,shapes.geometric}
\pgfdeclarelayer{signal}
\pgfsetlayers{signal,main}
\tikzstyle{printersafe}=[segment amplitude=0 pt]
\usetikzlibrary{arrows}
\newcounter{cntr}
\usetikzlibrary{calc}
\usetikzlibrary{decorations.pathreplacing,decorations.markings,shapes.geometric}
\tikzset{naming/.style={align=center}}
\tikzset{antenna/.style={insert path={-- coordinate (ant#1) ++(0,0.5) -- +(135:0.5) + (0,0) -- +(45:0.5)}}}
\tikzset{station/.style={naming,draw,shape=dart,shape border rotate=90, minimum width=20mm, minimum height=20mm,outer sep=0pt,inner
    sep=3pt}}
\tikzset{mobile/.style={naming,draw,shape=rectangle,minimum width=12mm,minimum height=6mm, outer sep=0pt,inner sep=3pt}}
\tikzset{radiation/.style={{decorate,decoration={expanding waves,angle=90,segment length=6pt}}}}

\tikzset{
  every pin/.style={rectangle,rounded corners=3pt,font=\footnotesize},
  small dot/.style={fill=black,circle,scale=0.5}
}

\tikzset{
  invisible/.style={opacity=0},
  visible on/.style={alt={#1{}{invisible}}},
  alt/.code args={<#1>#2#3}{%
    \alt<#1>{\pgfkeysalso{#2}}{\pgfkeysalso{#3}} 
  },
}

\tikzset{pics/.cd,
  SBS/.style={code={
      \begin{scope}[local bounding box=#1]
        \fill [pic actions/.try] (-1,0) -- (-1/2,3) -- (1/2, 3) -- (1,0) -- cycle;
        \fill [pic actions/.try] (-1/16,2) rectangle (1/16,4);
        \fill [pic actions/.try] (0,4) circle [radius=1/4];
        \foreach \i in {-1,1}
        \fill [shift=(90:4), xscale=\i]
        \foreach \r in {1,3/2,2}{
          (-45:\r) arc (-45:45:\r) -- (45:\r-1/10)
          arc(45:-45:\r-1/10) -- cycle
        };
      \end{scope}
    }},
  MBS/.style={code={
      \begin{scope}[local bounding box=#1]
        \fill [pic actions/.try] (-1,0) -- (-1/2,3) -- (1/2, 3) -- (1,0) -- cycle;
        \fill [pic actions/.try] (-1/16,2) rectangle (1/16,4);
        \fill [pic actions/.try] (0,4) circle [radius=1/4];
        \foreach \i in {-1,1}
        \fill [shift=(90:4), xscale=\i]
        \foreach \r in {1,3/2,2}{
          (-45:\r) arc (-45:45:\r) -- (45:\r-1/10)
          arc(45:-45:\r-1/10) -- cycle
        };
      \end{scope}
    }},
  SU/.style={code={
      \begin{scope}[local bounding box=#1]
        \fill [even odd rule, pic actions/.try]
        (-1,-5/2) -- (-1,-1/8) -- (1,-1/8) -- (1,-5/2)
        arc (360:180:1 and 1/4) -- cycle (-1,5/2) -- (-1,1/8) -- (1,1/8) -- (1,5/2)
        arc (0:180:1 and 1/4) -- cycle (-3/4, 9/4) -- (-3/4, 3/8) -- (3/4, 3/8) -- (3/4, 9/4)
        arc (0:180:3/4 and 1/8)-- cycle
        \foreach \i in {-1,0,1}{\foreach \j in {1,2,3}{
            (-\i*1/2-3/16,-\j/2-3/4) rectangle ++(3/8, 3/8)
          }
        }
        (-1/2,-3/4) rectangle (1/2, -1/4);
      \end{scope}
    }},
  MU/.style={code={
      \begin{scope}[local bounding box=#1]
        \fill [even odd rule, pic actions/.try]
        (-1,-5/2) -- (-1,-1/8) -- (1,-1/8) -- (1,-5/2)
        arc (360:180:1 and 1/4) -- cycle (-1,5/2) -- (-1,1/8) -- (1,1/8) -- (1,5/2)
        arc (0:180:1 and 1/4) -- cycle (-3/4, 9/4) -- (-3/4, 3/8) -- (3/4, 3/8) -- (3/4, 9/4) arc (0:180:3/4 and 1/8)-- cycle
        \foreach \i in {-1,0,1}{
          \foreach \j in {1,2,3}{
            (-\i*1/2-3/16,-\j/2-3/4) rectangle ++(3/8, 3/8)
          }
        }
        (-1/2,-3/4) rectangle (1/2, -1/4);
      \end{scope}
    }},
  SIGNAL/.style={code={
      \begin{scope}[local bounding box=#1]
        \fill [pic actions/.try]
        (0,-3) -- (-1,1/2) -- (1/8,1/4) -- (0,3) -- (1,-1/2) -- (-1/8,-1/4) -- cycle;
      \end{scope}
    }},
  queuei/.style={code={
      \begin{scope}
        \stepcounter{cntr}
        \node[inner sep=0pt, outer sep=0pt,draw,rectangle split,rectangle split horizontal,minimum height=0.5cm,rectangle split parts=3]
        (queue-\thecntr) [pic actions] {};
        \draw
        (queue-\thecntr.north west) -- ++(-0.2cm,0)
        (queue-\thecntr.south west) -- ++(-0.2cm,0);
        \node[above] at ([xshift=-0.5cm]queue-\thecntr.north)
        {$Q_#1$};
      \end{scope}
    }}
}
\colorlet{sky blue}{blue!60!cyan!75!black}
\colorlet{dark blue}{blue!50!cyan}
\colorlet{chameleon}{olive!75!green}
\tikzset{signal/.style={->,draw=black, line width=0.05em, dashed,printersafe}}

\newsavebox{\mybox}
\pgfplotsset{compat=newest}
\begin{document}

\title{Deep Deterministic Policy Gradient to Minimize the Age of Information in Cellular V2X Communications}

\author{Zoubeir~Mlika,~\IEEEmembership{Member,~IEEE}, and Soumaya~Cherkaoui,~\IEEEmembership{Senior~Member,~IEEE}}%

\maketitle

\begin{abstract}
  This paper studies the problem of minimizing the age of information (AoI) in cellular vehicle-to-everything communications. To provide minimal AoI and high reliability for vehicles' safety information, non-orthogonal multiple access is exploited. We reformulate a resource allocation problem that involves half-duplex transceiver selection, broadcast coverage optimization, power allocation, and resource block (RB) scheduling. First, to obtain the optimal solution, we formulate the problem as a mixed-integer nonlinear programming problem and then study its NP-hardness. The negative result of NP-hardness motivates us to design efficient sub-optimal solutions. Consequently, we model the problem as a single-agent Markov decision process (MDP). The MDP model helps in solving the problem efficiently using fingerprint deep reinforcement learning (DRL) techniques such as deep-Q-network (DQN) methods. Nevertheless, applying DQN is not straightforward due to the curse of dimensionality implied by the large and mixed action space that contains discrete RB scheduling decisions and continuous power and coverage optimization decisions. Therefore, to solve this mixed discrete/continuous problem efficiently simply and elegantly, we propose a decomposition technique that consists of first solving the discrete subproblem using a matching algorithm based on state-of-the-art stable roommate matching and then solving the continuous subproblem using DRL algorithm that is based on deep deterministic policy gradient (\texttt{DDPG}). We validate our proposed method through Monte Carlo simulations where we show that the decomposed matching and DRL algorithm successfully minimizes the AoI and achieves almost $66\%$ performance gain compared to the best benchmarks for various vehicles' speeds, transmission power, or packet sizes. Further, we prove the existence of an optimal value of broadcast coverage at which the learning algorithm provides the optimal AoI.
\end{abstract}


\newcommand{\describeContent}[1]{%
\begingroup%
\let\thefootnote\relax%
\footnotetext{#1}%
\endgroup%
}

\IEEEpeerreviewmaketitle


\section{Introduction}\label{sec:intro}
The vehicle-to-everything (V2X) ecosystem is one of the key verticals in fifth-generation (5G) networks~\cite{vannithamby20205g}. It enables vehicles to communicate with each other, with road infrastructure, and with other road users. It is an instrumental enabler for smart cities and intelligent transportation systems (ITS)~\cite{9186820}. Future generations of cellular networks (5G and beyond) are expected to meet the connectivity requirements of the automotive industry, including the V2X ecosystem, by providing the necessary wireless technology.

Many use cases for the automotive vertical can be supported, as outlined in the third generation partnership project (3GPP) Release 15 (3GPP Rel-15), including fleet management and infotainment~\cite{7917274}. In Rel-16, other advanced services are investigated by new radio (NR) that are grouped into four categories: extended sensors, vehicle platooning, advanced driving, and remote driving~\cite{3gpp.22.886}. 

We can broadly organize the different use cases into three groups of services: (i) safety services, (ii) non-safety services, and (iii) infotainment services. The main use case of V2X communication belongs to the first group of services, which aims to improve road safety through efficient and timely information exchange between vehicles, with road users, and/or with road infrastructure, to avoid road accidents. Important use cases of the safety services include vehicle detection, vehicle platooning, remote driving, to name a few. 

Safety services often require ultra-reliable, low latency communication (URLLC) and timely information exchange. For example, for a platoon of vehicles with a minimum distance of 2 meters between vehicles, the end-to-end latency requirements should be about 25 milliseconds to support a speed of 100 km/h. In contrast, for a remote driving application, the end-to-end latency should not exceed 5 milliseconds~\cite{vannithamby20205g}. In addition to the delay or latency metrics, the age of information (AoI)~\cite{8006590} plays a key role in V2X communication~\cite{10.1145/3331054.3331547} to ensure timely information exchange. In a V2X ecosystem, to provide safety services, it is necessary that the information exchanged between vehicles is fresh. The freshness of information can be measured by the AoI that plays a paramount importance in vehicular networks. 

AoI captures delay from an application layer perspective and is defined as the time between the generation of information and its successful delivery~\cite{8006590}. AoI includes not only delay but also inter-delivery time, which is a more important and vital performance metric for safety services in vehicular networks. The inter-delivery time is indeed taken into account by AoI since the latter keeps increasing as long as the information has not been updated. 

The objective of this paper is to develop algorithmic solutions that successfully minimize the average AoI in 5G vehicular networks. To do so and thus to support V2X communication with minimal AoI while meeting the corresponding stringent safety service requirements, cellular network infrastructures such as the widely deployed long-term evolution (LTE) have been considered. In 3GPP Rel-16, NR cellular V2X communication can support a direct transmission mode called sidelink~\cite{9088326,vannithamby20205g}, and a transmission mode over the cellular network, which together provides seamless connectivity to vehicles through a unified radio. 

Resource allocation for V2X side-link communication can be performed in Mode 1 or Mode 2~\cite{vannithamby20205g}. The first mode is centralized where the roadside units (RSU) schedule the resources for V2X side-link communication, while the second mode is distributed where vehicles autonomously allocate network-configured resources for V2X side-link communication. Thanks to the promising solution of the cellular-based V2X network that provides high cellular coverage and very low latency even in high mobility scenarios~\cite{7974737}, we adopt, in this paper, the centralized resource allocation approach. Despite all the benefits that the cellular-based V2X network can provide, optimizing AoI to support safety services in a V2X ecosystem is a challenging problem from an algorithmic perspective. 

Due to the broadcast nature of vehicular communications, it is difficult to guarantee a minimum AoI for security packets and the solution to this problem has not been well answered in the literature, especially from optimization and algorithmic point of view. Minimizing AoI is even more complex when the orthogonal multiple access (OMA) technique is used because OMA prevents multiple vehicles from broadcasting over the same resource due to interference~\cite{7974737}. Therefore, for better AoI and better spectral efficiency, the power-domain non-OMA (NOMA) technique~\cite{6692652} is proposed as a potential solution for beyond 5G networks~\cite{7974737}. Thus, multiple vehicles (a NOMA group), can broadcast their safety information simultaneously over the same resource~\cite{6868214}. To decode the information of each vehicle, various multi-user detection algorithms have been proposed, including the well-known successive interference cancellation (SIC). Despite the difficulty of performing decoding in a large NOMA group (due to large processing times), the powerful computation capabilities of advanced integrated circuit chips allow the complexity of SIC to be reduced and the delays incurred to be disregarded~\cite{7974737}. 

The use of NOMA in vehicular networks to optimize AoI remains challenging due to (i) broadcasting and (ii) half-duplex\footnote{As noted in~\cite{7974737}, for simplicity and due to the implementation problems of full-duplex, all vehicles operate in the half-duplex mode.} modes of operation. Due to the half-duplex, when multiple vehicles broadcast their information in the same resource block (RB), then no vehicle is capable of receiving the correct information from within its coverage area. Further, broadcasting under the NOMA technique brings even more severe interference to other vehicles, pedestrians, and infrastructure due to limited spectrum resources. To solve the above two issues, wise RBs scheduling and power allocation (to correctly decode the received information in a non-orthogonal manner, the transmission power of each transmitter vehicle should be carefully allocated) and careful broadcast coverage range optimization of each vehicle are required. All of the above decisions should be optimized while guaranteeing up-to-date information transmission to minimize the AoI of the vehicles. Therefore, it is necessary to rethink the state-pf-the-art resource allocation and propose new algorithmic techniques to optimize the vehicles' coverage and resources to minimize the AoI in a NOMA-based vehicular network while ensuring different constraints such as i) a half-duplex vehicle does not transmit and receive safety information over the same RB, ii) a maximum transmission power. Note that the optimization of vehicle coverage is not well studied in the literature, although it is argued that grouping vehicles into clusters is particularly important for reducing interference and thus minimizes AoI~\cite{5984917,8937801}. Therefore, the considered problem statement is the minimization of AoI using the NOMA technique in half-duplex vehicular networks while deciding the coverage areas of vehicles, scheduling their safety information, and allocating the transmission power of each NOMA group.

\subsection{Related Works}\label{sec:sota}

In~\cite{7974737}, B. Di \textit{et al.} considered the resource allocation problem in V2X communication networks using NOMA. Specifically, the problem is formulated using mathematical programming to maximize the number of successfully decoded signals, which involved three decision-making: (i) transceiver selection in a half-duplex scenario, (ii) frequency allocation, and (iii) power allocation. The authors first have shown the NP-hardness of the problem by reducing the vertex cover problem to a special case of their problem, then, using matching theory, they proposed centralized semi-persistent scheduling (SPS)-based approach, as in Mode 1, to handle the first two decision making and designed a rotation-based matching algorithm using the multidimensional stable roommate problem. Next, they proposed an iterative power control scheme based on a distributed implementation. Our work can be seen as an extension of this work but B. Di \textit{et al.} did not optimize AoI and their complexity analysis does not apply to our case.

In~\cite{9205620}, A. H. Sodhro \textit{et al.} proposed a machine learning-driven mobility management method for industrial network in box (NIB) applications to intelligently allocate the resources with high energy efficiency. The authors also proposed a novel architecture of 6G-based intelligent quality-of-experience (QoE) and quality-of-service (QoS) optimization in industrial NIB. A 6G NIB framework in association with LTE networks is further studied and an energy-efficient use case for 6G industrial NIB is presented. The authors used artificial intelligence and mathematical optimization to derive their framework. The authors showed, using real-time datasets, MATLAB and CVXOPT, that the proposed mobility management method provides high energy efficiency and better QoE and QoS in 6G-based industrial NIB.

In~\cite{9199859}, A. H. Sodhro \textit{et al.} proposed a joint cross-layer approach for the development of green, sustainable, reliable, and smart 5G-based ITS. The cross-layer based reliability optimization approach combines four layers, including physical layer (energy efficiency), medium access control layer (battery-aware), network layer (reliability), and application layer (signaling and transmission), by adopting modulation level monitoring, node's duty cycle, routing/processing, and particular application selection. Energy and reliability optimization problems were formulated and real-time datasets were used in MATLAB to evaluate the performance.

In~\cite{9085258}, A. H. Sodhro \textit{et al.} studied the problem of reliable channel modeling in fog computing vehicular networks and they proposed reliable and delay-tolerant channel model with better QoS and they optimized QoS in terms of mobility, reliability, and packet loss. For fog computing intra-vehicle networks, the authors proposed, based on artificial intelligence, a reliable and interference-free mobility management algorithm to provide efficient communication, computation, cooperation, and storage. Next, for fog computing inter-vehicle networks, they proposed a reliable and efficient multi-layer framework. Real-time datasets and the convex optimization framework of MATLAB, as well as the ANOVA platform, were used to evaluate the proposed solution. Further, the authors discussed some challenges and possible solutions for intercity vehicular communications including mobility management, interference mitigation, power control, and routing.

In~\cite{9014535} A. H. Sodhro \textit{et al.} studied the problem of maximizing the reliability and connectivity and minimizing the packet loss in software-defined Internet of vehicles networks. The authors first proposed, based on the key features of the wireless channel, a system model for reliability and connectivity optimization in software-defined Internet of vehicles. Then, a stable and scalable link optimization algorithm, that adopts the key characteristics of the reliability and connectivity optimization in vehicular networks, is proposed and compared to baseline methods. Next, a novel vehicular connectivity, and reliability framework is proposed that comprises different steps including vehicular connectivity to the Internet, collecting and monitoring the vehicle’s data, and connecting the vehicles for reliable, and connectivity optimized communication. Further, a smart city use case was studied.

Despite the interesting works of A. H. Sodhro \textit{et al.} in~\cite{9205620,9199859,9085258,9014535}, the authors did not optimize the AoI and thus the formulated problem in our work is mainly different.

In~\cite{8937801}, M. K. Abdel-Aziz \textit{et al.} studied the problem of power minimization in vehicle-to-vehicle (V2V) communication networks by considering the tail distribution of AoI. The authors used extreme value theory and queuing theory to study extreme AoI events that go beyond the average AoI analysis. It is assumed that the vehicles are grouped by an RSU into clusters based on their geographic locations. The problem is formulated as a time-averaged power minimization with probabilistic AoI constraints in both the deterministic and Markovian arrival cases and is solved using Lyapunov stochastic optimization. The authors formulated their problem differently from ours and did not focus on optimizing the half-duplex transceiver selection, coverage optimization, and resource allocation. A Manhattan mobility model was adopted in the Monte Carlo simulations.

The minimization of the AoI is studied in energy harvesting cellular networks~\cite{9128850} where the authors studied the problem of user scheduling and user association and showed the existence of AoI-fairness tradeoff. Note that the authors did not study the minimization of the AoI in vehicular networks where the optimization of the coverage areas of the vehicles is an important problem.



In~\cite{8954939}, X. Chen \textit{et al.} considered a Manhattan grid V2V network and formulated a single-agent Markov decision process (MDP) that aims to improve the expected long-term performance of all vehicles. An RSU is assumed to make decisions, as in the Mode 1 case discussed earlier, for vehicles by allocating frequency bands and scheduling packets over time. It is assumed that the RSU can choose the groups of vehicles based on geographic locations. Due to the high mobility and variation of vehicle traffic, the resource allocation decision space can grow exponentially. Therefore, the authors proposed an MDP decomposition technique using the Q-function decomposition. Using long short-term memory from the recurrent neural network and deep reinforcement learning (DRL), a proactive AoI-aware algorithm is proposed to optimize the AoI in a decentralized manner. However, the constraints of the studied problem are different from our constraints such as the half-duplex transceiver selection, and coverage optimization. Evaluations were conducted using the Python programming language with TensorFlow framework where the algorithms were trained in an offline manner.

\begin{figure*}
  \centering
  \includegraphics[width=0.7\textwidth]{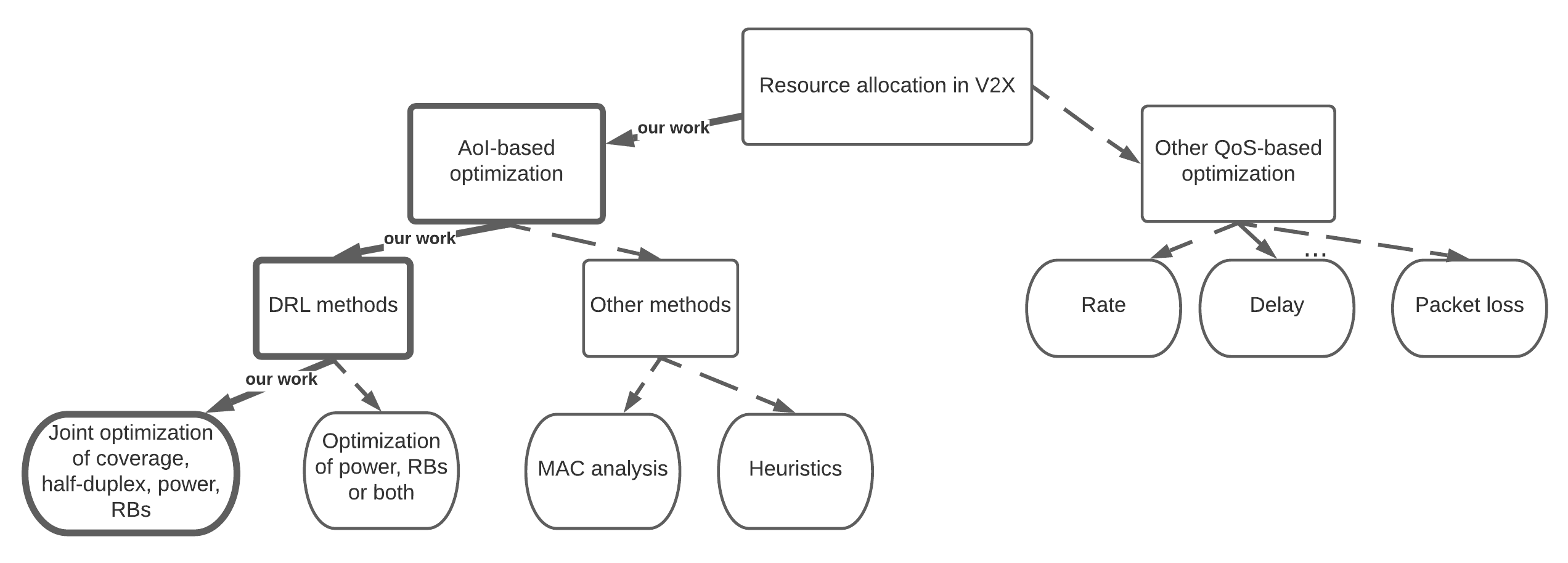}
\caption{A flow chart of the position of our work in comparison to the related works. Our work is represented with solid arrows whereas other related works are represented with dashed arrows.}
\label{flow}
\end{figure*}

In~\cite{9195789}, M. Samir \textit{et al.} studied AoI minimization in a network of unmanned aerial vehicles (UAVs). The problem formulated is in the form of a mixed-integer non-linear program that aims to optimize the UAVs' trajectories as well as scheduling decisions while guaranteeing minimum throughput constraints. By modeling the problem as an MDP, the authors proposed a DRL approach based on the deep deterministic policy gradient (\texttt{DDPG}) algorithm. A complexity analysis of the proposed solution was also given. However, the paper did not consider the coverage optimization and the half-duplex transmission (transceiver selection) of the vehicles. The authors used Monte Carlo simulations and trained their proposed solutions using tensor processing units.

In~\cite{9214855}, F. Peng \textit{et al.} studied the AoI performance of SPS in a V2X communication network. Due to packet collisions and half-duplex effects of the SPS-based MAC layer protocol, the AoI is shown to deteriorate. The authors proposed a collaborative approach based on piggyback mechanisms in which vehicles act cooperatively to inform each other on potential collisions. However, the paper only considers the minimization of AoI from a MAC layer perspective, and no algorithmic solution was proposed.Theoretical performance analysis in the scenarios of both static and dynamic vehicular traffic flows was presented and its convergence was studied. Monte Carlo simulations were performed in a freeway scenario.

In~\cite{8792382}, L. Liang \textit{et al.} studied the problem of spectrum sharing and power allocation in vehicular networks. The objective is formulated, using MDP, to maximize the total link throughput between the vehicle and the infrastructure while guaranteeing the URLLC requirements of V2V links. The authors proposed, using proper reward design, a multi-agent MDP that is solved in a distributed manner using fingerprint-based DQN methods. The proposed solution was simulated using Monte Carlo methods and was shown to allow the vehicles to learn from interactions with the vehicular environment and figure out a clever cooperative strategy in a distributed way. However, the paper did not optimize the AoI.

The application of DQN techniques to solve the resource allocation problem in wireless networks with applications to vehicular networks is discussed in~\cite{8943940}. Also, in~\cite{mlika2021network,9318243} the authors studied the problem of resource allocation in vehicular networks with network slicing and proposed efficient deep reinforcement learning solutions based on DQN. Further, a deep reinforcement learning algorithm is proposed in~\cite{9524882} to solve the problem of service migration in multi-access edge computing (MEC)-enabled vehicular networks. However, fingerprint-based DQN is not directly applicable to our problem due to the curse of dimensionality and a more continuous control based reinforcement learning approach is needed.

We identify important research gaps that were not solved by previous works. In particular, previous work (i) has optimized only access latency; however, we believe that optimizing the AoI metric is more important in vehicular networks due to the need to provide up-to-date information for safety-critical applications. Or, (ii) the works have focused on the MAC perspective of AoI in vehicular networks, but we believe that considering AoI optimization from an algorithmic perspective is more relevant to solving the intrinsic difficulty of the problem. Or, (iii) the works did not optimize the coverage along with the half-duplex transceiver selection to minimize the AoI.

We show in Fig.~\ref{flow} the position of our work in comparison to other related works that include resource allocation based on AoI using DRL methods that optimize jointly the coverage optimization, the half-duplex transceiver selection, power allocation, and RBs scheduling.

In this paper, we fill these research gaps by studying a V2X ecosystem to minimize the average AoI while applying the NOMA technique. NOMA is used to wisely use the limited wireless resources by different vehicles in a non-orthogonal manner. We solve the joint problem of half-duplex transceiver selection, coverage optimization, and wireless resource allocation (i.e., time slots scheduling and power allocation) according to the following methodology. Specifically, the joint problem involves complex mixed decisions that include discrete decisions (half-duplex selection and resource block allocation) and continuous decisions (coverage optimization and NOMA power allocation). To solve this challenging problem efficiently, the discrete-based decisions problem is modeled as a matching problem according to~\cite{7974737} and solved using state-of-the-art method~\cite{7974737} (though~\cite{7974737} did not optimize AoI). However, the continuous-based decisions problem, which cannot be discretized, is modeled as a Markov decision process (MDP) and solved using deep reinforcement learning (DRL) methods based on deep deterministic policy gradient (\texttt{DDPG}). To obtain an optimal solution to this joint problem using off-the-shelf solvers (for small-scale networks at least due to the high computational complexity of solver-based solutions), it is formulated as a mixed-integer non-linear program (MINLP). Next, we provide a rigorous NP-hardness analysis of the problem. Note that, in general, a particular problem is MINLP does not necessarily imply its NP-hardness. The most directly related work to ours is~\cite{7974737} in which the authors did not consider AoI optimization, only proposed simple heuristic and greedy methods, and studied the NP-hardness of the problem only in the case of frequency allocation, i.e., when the system involved single frequency channel, it is not obvious to generalize the NP-hardness proof presented in~\cite{7974737}.

Note that the modeling of the entire problem (discrete and continuous) using MDP and DRL is also possible but is more involved. This is regarded as a limitation in our work and should be considered for future works as discussed in the conclusion. Our contributions are listed below:
\begin{itemize}
	\item We model the AoI minimization problem as an MINLP;
	\item We rigorously prove its NP-hardness;
	\item We model the AoI minimization problem as a single-agent Markov decision process.
	\item We solve the AoI minimization problem using a decomposition technique that first solves a matching subproblem and then solves an MDP problem. The matching algorithm finds the discrete actions by assigning the RBs to vehicles using state-of-the-art stable roommate matching techniques. Next, the continuous control DRL algorithm uses the deep deterministic policy gradient (\texttt{DDPG}) to find the transmission power levels and broadcast coverage ranges. 
	\item We show through extensive simulations that the decomposition-based matching and DRL algorithm successfully minimizes the AoI and achieves better AoI compared to benchmarks.
\end{itemize}


\subsection{Notations}
The transmission link between $i$ and $j$ is denoted in subscript as $i\rightarrow j$. We use boldface letters to denote multidimensional vectors, i.e., $\mathbf{x}\coloneq[x_{i\rightarrow j}^{(t)}]$ is a 3D matrix with indices $i$, $j$, and $t$, and $\mathbf{p}\coloneq[p_{i}^{(s)}]$ is a 2D matrix of indices $i$ and $s$. Calligraphic letters denote sets, e.g., the set of the first $n$ positive integers is denoted by $\mathscr{A}=\{1,2,\ldots,n\}$. The calligraphic letter with parentheses $\mathscr{O}(\cdot)$ denotes the big-O notation. In general, the superscript denotes time index and is giving in parentheses (e.g., $x^{(s,t)}$ denotes the variable $x$ at time instant $s$ of the period $t$) to remove remove any confusion about exponentiation. The binary logarithm of a variable $x$ is denoted by $\lg(x)$. Other notations are defined when used.

\section{Model}\label{sec:mod}
\subsection{System Model}
    We consider a cellular-based V2X communication network in which there are $m$ half-duplex vehicles denoted by the set $\mathscr{V}$ and a road-side unit (RSU). Time is discrete and is divided into time slots of duration $\tau$ seconds each ($\tau$ is on the order of a few milliseconds). Each set of $k$ consecutive time slots is denoted by $\mathscr{S}$ and forms what we call a transmission period. We consider a time window of $n$ transmission periods denoted by the set $\mathscr{T}$. In this work, for simplicity, we consider a single carrier system model in which a time slot simply represents a resource block (RB). The multi-carrier system model is a simple extension that does not affect our theoretical analysis of the problem. Each RB has a bandwidth of $\omega$ Hertz.
    
    At the beginning of each transmission period $t$, each vehicle $i$ generates its safety information, or simply, a safety packet of size $\zeta_i^{(t)}$ bits, and must transmit it to all other vehicles using broadcasting\footnote{When there is no confusion, we omit the transmission period index $t$.}. The vehicles communicate with each other via side-link communication where the RSU acts as a central controller that makes decisions for the vehicles. Multiple vehicles can transmit on the same RB using NOMA. Indeed, the use of NOMA improves spectral efficiency and ensures low latency and high reliability for vehicular applications~\cite{7974737}. However, due to half-duplex communication and broadcasting, a vehicle cannot transmit and receive on the same RB. 
    
\begin{figure*}
  \centering
  \includegraphics[width=0.7\textwidth]{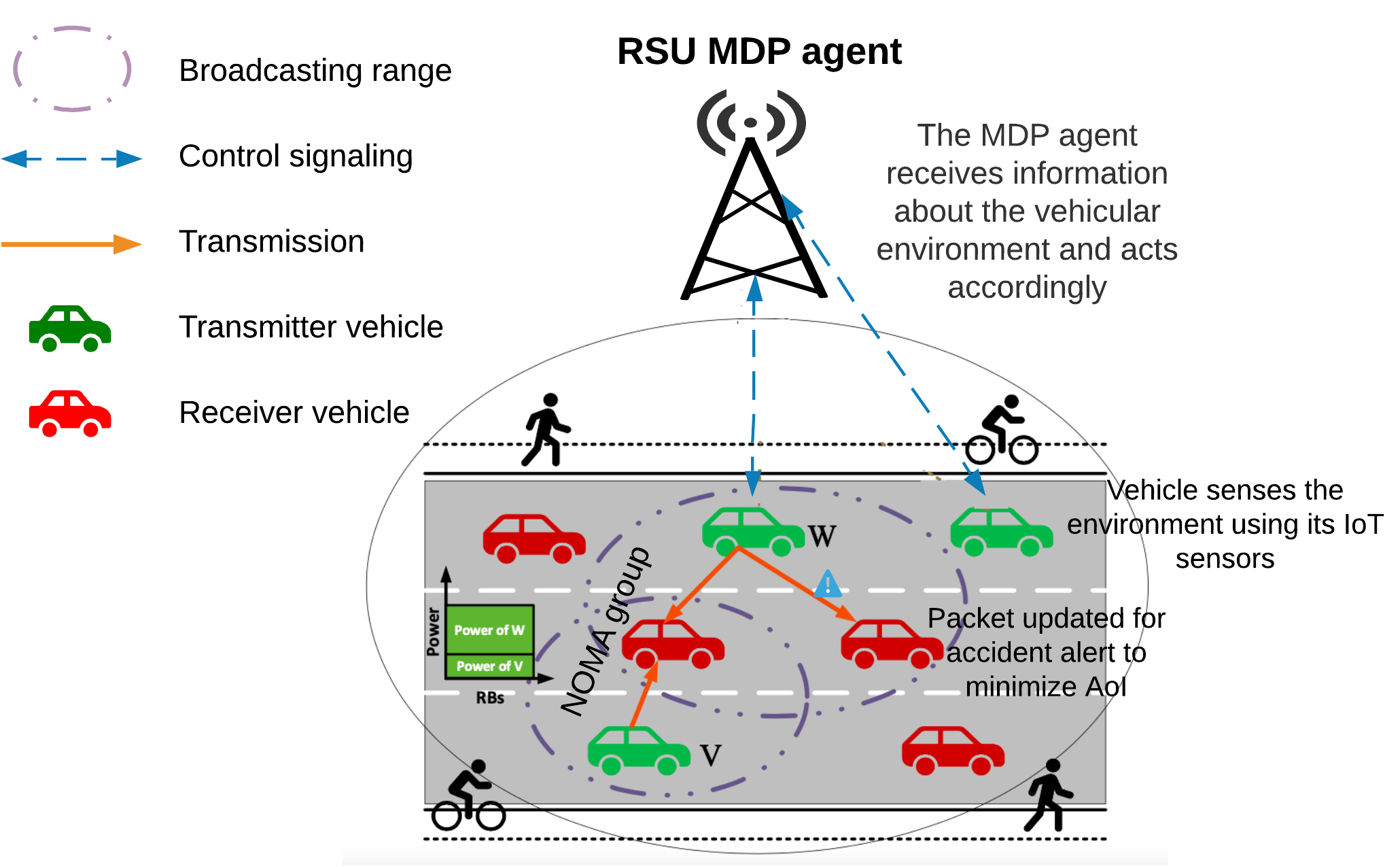}
\caption{An example of the considered V2X ecosystem with one RSU. There are two transmitter vehicles (V and W) and four-receiver vehicles. Due to the coverage area of V and W, only two receiver vehicles are receiving information. Vehicles V and W use NOMA to communicate to the same red receiving vehicle. Minimizing the AoI is can prevent an accident from happening by acting proactively due to frequently updated information.}
\label{sysmod1}
\end{figure*}

    In addition, the transmission power must be carefully allocated to achieve the promised gains of NOMA. In summary, once each vehicle $i\in\mathscr{V}$ has generated its safety information, it must make four decisions: (i) choose its role as transmitter or receiver, (ii) choose a set of receiving vehicles by choosing its coverage range (in the case of a transmitter), (iii) assign a set of time slots during which it will transmit, and (iv) allocate, for each chosen time slot $s$, a transmission power $p_i^{(s)}$ that must not exceed a maximum transmission power $\overline{p}_i$. In the considered transmission scenario, all decisions are made by the RSU in a centralized way as it is done in~\cite{7974737,8954939}.
    An example of our system model is given in Fig.~\ref{sysmod1}.

\subsection{Mathematical Model}
    The main objective is to improve the freshness of safety information in vehicular networks. Information freshness is measured using the well-known AoI metric. When vehicle $i\in\mathscr{V}$ transmits to vehicle $j\in\mathscr{V}$, the AoI of link $(i,j)$ up to the transmission period $t$ is denoted by $\Delta_{i\rightarrow j}^{(t)}$. We use the arrow notation here to denote the transmission direction between vehicles. Based on our discrete-time model and the periodic nature of safety information in vehicular networks, the AoI $\Delta_{i\rightarrow j}^{(t)}$ denotes the number of transmission periods that have elapsed up to the transmission period $t$ since the last successful transmission between $i$ and $j$. The smaller the $\Delta_{i\rightarrow j}^{(t)}$, the better the freshness of the information. Note that the use of discrete transmission periods as units of AoI is considered in~\cite{7852321,10.1145/3209582.3209602}. 
    
    To mathematically define $\Delta_{i\rightarrow j}^{(t)}$, we introduce the binary variable $x_{i\rightarrow j}^{(t)}$ to denote the successful transmission between vehicles $i$ and $j$. In other words, $x_{i\rightarrow j}^{(t)}=1$ if and only if vehicle $j$ successfully receives the safety information generated by vehicle $i$ during the transmission period $t$. That said, the AoI of the link $(i,j)$ can be recursively calculated as follows:
\begin{align}
	\label{aoi1}
    \Delta_{i\rightarrow j}^{(t)}\coloneq
	\begin{cases}
		1, & \text{if $x_{i\rightarrow j}^{(t-1)}=1$},\\
		1+\Delta_{i\rightarrow j}^{(t-1)}, & \text{otherwise.}
	\end{cases}
\end{align}
That is, the AoI of the safety packet $\zeta_i^{(t)}$ of the link $(i,j)$ has the following two properties: (i) it increases by one as long as it is not successfully received by $j$ in the previous transmission period, and (ii) it goes back to one when it is successfully received by $j$ in the previous transmission period. Thus, the AoI, $\Delta_{i\rightarrow j}^{(t)}$, can be simply given by the formula:
\begin{align}
	\label{aoi2}
	\Delta_{i\rightarrow j}^{(t)}=1+(1-x_{i\rightarrow j}^{(t-1)})\Delta_{i\rightarrow j}^{(t-1)}.
\end{align}
Here we see that when $x_{i\rightarrow j}^{(t-1)}=1$, meaning that the information was received in $t-1$ (and thus it is up-to-date), then the AoI is equal to 1 (the minimum value). However, when $x_{i\rightarrow j}^{(t-1)}=0$, meaning that the information was not received previously (and thus it is outdated), then the AoI increases by 1.

When vehicle $i\in\mathscr{V}$ communicates with vehicle $j\in\mathscr{V}$ at time slot $s$ during the transmission period $t$, the signal-to-interference-plus-noise ratio (SINR) is given by:
\begin{align}
	\label{sinr1}
	\gamma_{i\rightarrow j}^{(s,t)}\coloneq\dfrac{p_i^{(s)}g_{i\rightarrow j}^{(s,t)}}{1+I_{i\rightarrow j}^{(s,t)}}.
\end{align}
In~\eqref{sinr1}, $g_{i\rightarrow j}^{(s,t)}$ denotes the normalized power gain of the channel, i.e., $g_{i\rightarrow j}^{(s,t)}=|h_{i\rightarrow j}^{(s,t)}|^2/N_0$ where $h_{i\rightarrow j}^{(s,t)}$ is the channel coefficient which includes large and small scale fading and $N_0$ is the noise power over the considered bandwidth $\omega$. The interference term in the denominator, $I_{i\rightarrow j}^{(s,t)}$, is calculated according to the NOMA principle in which multiple vehicles (including $i$) transmit to vehicle $j$. Specifically, the interference $I_{i\rightarrow j}^{(s,t)}$ at the $j$th receiving vehicle comes from the transmitting vehicles (different from $i$) that are scheduled at time slot $s$ of the transmission period $t$ and transmit to vehicle $j$. Vehicle $j$ then applies SIC to decode the received information. It is well-known that the highest channel decoding order should be applied in this uplink-like NOMA scenario~\cite{8632657}. In other words, the interference $I_{i\rightarrow j}^{(s,t)}$ received by vehicle $j$ comes from all vehicles other than $i$ that have lower channel power gains. 

Let $\mathscr{I}_{i\rightarrow j}^{(s,t)}$ be the set of vehicles other than $i$ that transmit to vehicle $j$ at time slot $s$ of the transmission period $t$ such that $g_{i\rightarrow j}^{(s,t)}>g_{i'\rightarrow j}^{(s,t)}$ for all $i'\in\mathscr{I}_{i\rightarrow j}^{(s,t)}$, i.e., $\mathscr{I}_{i\rightarrow j}^{(s,t)}=\{i'\in\mathscr{V}\backslash\{i\}:g_{i\rightarrow j}^{(s,t)}>g_{i'\rightarrow j}^{(s,t)}\}$. Therefore, the interference term $I_{i\rightarrow j}^{(s,t)}$ is given mathematically as follows:
\begin{align}
	\label{interference1}
	I_{i\rightarrow j}^{(s,t)}\coloneq\sum_{i'\in\mathscr{I}_{i\rightarrow j}^{(s,t)}}p_{i'}^{(s)}g_{i'\rightarrow j}^{(s,t)}x_{i'\rightarrow j}^{(t)}.
\end{align}
For example, in Fig.~\ref{sysmod1} the set $\mathscr{I}_{V\rightarrow X}^{(s,t)}$ for a given pair $(s,t)$ is given by vehicle $W$ where $X$ denotes the red vehicle to which $V$ and $W$ are transmitting to.

Based on the SINR, the achievable rate (based on Shannon theorem) on link $(i,j)$ at time slot $s$ of transmission period $t$ can be obtained mathematically as follows:
\begin{align}
	\label{rate1}
	R_{i\rightarrow j}^{(s,t)}=\omega\tau\lg\bigl(1+\gamma_{i\rightarrow j}^{(s,t)}\bigr).	
\end{align}

The resource allocation problem considered is called the vehicular age minimization problem (VAMP). Its objective is to minimize the average AoI of all vehicles while associating vehicles with each other, scheduling transmission to appropriate RBs, and allocating transmission powers. To solve VAMP, first, we formulate it as a mixed-integer nonlinear program (MINLP). This formulation is useful for obtaining the optimal solution using off-the-shelf solvers. Then, we study its NP-hardness. Finally, we propose a single-agent MDP formulation that provides an online suboptimal solution using DRL techniques.

VAMP can be formulated as the following MINLP.
\begin{subequations}
	\label{mp1}
	\begin{align}
    	\minimize_{\mathbf{x},\mathbf{y},\bm{\Delta},\mathbf{p}}
		    &\quad \sum_{i\in\mathscr{V}}\sum_{j\in\mathscr{V}}\sum_{t\in\mathscr{T}}\Delta_{i\rightarrow j}^{(t)},\label{obj}\\
    	\subjto
		& \quad y_i^{(s,t)},x_{i\rightarrow j}^{(t)}\in\{0, 1\},\forall i,j,s,t,\label{c1a}\\
		& \quad \Delta_{i\rightarrow j}^{(t)}\in\{0,1,\ldots,n+1\},\forall i,j,t,\label{c1b}\\
		& \quad \Delta_{i\rightarrow j}^{(0)}=1,x_{i\rightarrow j}^{(0)}=1,y_i^{(s,0)}=1\forall i,j,\label{c3}\\
		& \quad \Delta_{i\rightarrow j}^{(t)}=1+(1-x_{i\rightarrow j}^{(t-1)})\Delta_{i\rightarrow j}^{(t-1)},\forall i,j,t,\label{c2}\\
		& \quad \Delta_{i\rightarrow i}^{(t)}=0,x_{i\rightarrow i}^{(t)}=0,\forall i,t,\label{c4}\\
		& \quad \sum_{s\in\mathscr{S}}y_i^{(s,t)}R_{i\rightarrow j}^{(s,t)}\geqslant\zeta_i^{(t)}x_{i\rightarrow j}^{(t)},\forall i,j,t,\label{c5}\\
		& \quad x_{i\rightarrow j}^{(t)}\leqslant\sum_{s\in\mathscr{S}}y_i^{(s,t)},\forall i,j,t,\label{c6}\\
		& \quad y_{i}^{(s,t)}\leqslant\sum_{j\in\mathscr{V}}x_{i\rightarrow j}^{(t)},\forall i,s,t,\label{c7}\\
		& \quad y_i^{(s,t)}x_{i\rightarrow j}^{(t)}+y_{j'}^{(s,t)}x_{j'\rightarrow i}^{(t)}\leqslant1,\forall i,j,j',s,t,\label{c8}\\
		& \quad y_i^{(s,t)}x_{i\rightarrow j}^{(t)}+y_{j}^{(s,t)}x_{j\rightarrow i'}^{(t)}\leqslant1,\forall i,i',j,s,t,\label{c9}\\
		& \quad \sum_{i\in\mathscr{V}}y_i^{(s,t)}\leqslant\overline{m},\forall s,t,\label{c10}\\
		& \quad 0\leqslant p_i^{(s)}\leqslant\overline{p}_i,\forall i,s,\label{c11}
	\end{align}
\end{subequations}

In~\eqref{mp1}, the objective function given in~\eqref{obj} calculates the overall AoI of each pair of vehicles $(i,j)$ according to~\eqref{aoi2}. Constraints~\eqref{c1a}-\eqref{c4} illustrate the optimization variables, their initialization , and the evolution of the AoI. Here, $\Delta_{i\rightarrow j}^{(t)}$ represents the AoI of link $(i,j)$ up to transmission period $t$ and is an integer variable whose lower bound is $0$ and upper bound is $n+1$ (see~\eqref{aoi1}). The variable $y_{i}^{(s,t)}$ is binary that is equal to 1 if and only if vehicle $i$ is scheduled to transmit at time slot $s$ of transmission period $t$. The variable $x_{i\rightarrow j}^{(t)}=1$ if and only if vehicle $j$ successfully receives the safety information generated by vehicle $i$ during the transmission period $t$. The initialization of variables prior to the first transmission period is given in constraints~\eqref{c3}. The evolution of AoI is given by constraints~\eqref{c2}. Constraints~\eqref{c4} indicate that vehicle $i$ cannot transmit to itself and therefore must have an AoI equal to $0$ and $x_{i\rightarrow i}^{(t)}=0$.

Constraints~\eqref{c5} guarantee that if vehicle $i$ transmits to vehicle $j$ during transmission period $t$, then the total achievable throughput on the allocated RBs is at least the corresponding packet size. The relationship between the variables $x_{i\rightarrow j}^{(t)}$ and $y_i^{(s,t)}$ is given in constraints~\eqref{c6} and~\eqref{c7}. That is, (i) if vehicle $i$ successfully delivers its safety packet to vehicle $j$ during transmission period $t$, then there must exist at least one time slot $s$ at which $i$ has been scheduled, and (ii) if vehicle $i$ has never successfully delivered its safety packet to any vehicle during transmission period $t$, then it must not be scheduled at any time slot $s$. Due to the NOMA technique and half-duplex communication, constraints~\eqref{c8} ensure that vehicle $i$ cannot transmit to the vehicle $j$ and receive from vehicle $j'$ (including $j$) in the same time slot. Similarly, constraints~\eqref{c9} guarantee that vehicle $j$ cannot receive from vehicle $i$ and transmit to vehicle $i'$ (including $i$) in the same time slot. Constraints~\eqref{c10} limit the number of vehicles scheduled at time slot $s$ to $\overline{m}$ to reduce the decoding complexity in SIC. Finally, constraints~\eqref{c11} ensure that the transmission power of vehicle $i$ is nonnegative and does not exceed its maximum value.

Clearly, VAMP as formulated in~\eqref{mp1} is nonlinear. Note that, although constraints~\eqref{c2},~\eqref{c8}, and~\eqref{c9} can be easily linearized by introducing auxiliary binary variables, constraints~\eqref{c5} cannot be linearized due to the logarithm in~\eqref{rate1} as well as the expression of SINR and interference~\eqref{sinr1} and~\eqref{interference1}, respectively. Therefore, it is very challenging to solve VAMP optimally in polynomial-time. This conclusion is general and is often found in the literature. 
To justify this conclusion rigorously, we provide formal proof that VAMP is indeed NP-hard even for a restricted version. As mentioned earlier, a similar problem was studied in~\cite{7974737} that considered a single transmission period and optimized a different objective function. The authors studied the NP-hardness of the problem and showed by a restriction that the formulated problem is NP-hard. However, their proof is only valid for multi-carrier vehicular networks because they showed that the graph coloring problem is reducible to the subcarrier allocation problem. Therefore, the question of whether the single carrier problem as formulated in~\eqref{mp1} is NP-hard or not remains open. 

\subsection{NP-hardness}
In this subsection, we consider a restricted version of VAMP in which there are only two transmission periods and the OMA technique is used (i.e., the maximum number of vehicles that can be grouped together is $\overline{m}=1$). The problem in the NOMA case is more general and therefore more difficult than this restricted case. For this restricted case, to minimize the AoI, one must maximize the number of successfully received safety packets in the first transmission period. Indeed, as shown in~\eqref{aoi1}, the AoI up to the first transmission period is constant but the AoI up to the second transmission period depends on the number of packets successfully received in the first transmission period. The restricted version of VAMP will be called the VAMP$^\prime$. We prove the following result:
\begin{lemma}
	\label{lemma1}
    VAMP$^\prime$ is NP-hard.
\end{lemma}
\begin{IEEEproof}
    Once we show that VAMP$^\prime$ is NP-hard, then, by restriction~\cite{Garey:1979}, VAMP is NP-hard too. We reduce, in polynomial-time, the maximum independent set (MIS) problem~\cite{Garey:1979} to VAMP$^\prime$.

    Given a graph with $\ell$ vertices as an instance of MIS, we construct, in polynomial-time, an instance of VAMP$^\prime$ such that MIS is solved if and only if VAMP$^\prime$ is solved. The vertices of the graph correspond to the vehicles while its edges correspond to the time slots in the second transmission period. An additional vehicle $i^*$ is created. Thus, we create a total of $\ell+1$ vehicles denoted by $\{1,2,\ldots,\ell\}\cup\{i^*\}$. We consider a simple star network scenario (an uplink-like scenario) in which all vehicles in $\{1,2,\ldots,\ell\}$ wish to transmit to $i^*$, i.e., we can simply choose the channel gain between the vehicles in $\{1,2,\ldots,\ell\}$ to be very small and the transmission power of vehicle $i^*$ to be zero. Unless otherwise specified, the other network parameters (e.g., transmission power, noise power, etc.) are normalized to one. It remains to construct the channel gains between vehicle $i$ in $\{1,2,\ldots,\ell\}$ and vehicle $i^*$. The channel coefficient of each link $(i,i^*)$ at time slot $s$ is set to $1$ if $s$ is incident to vertex $i$ and is set to $0$ otherwise. In this way, the achievable rate between vehicle $i$ and $i^*$ at a time slot is either $1$ or $0$. The size of the safety packet, $\zeta_i$, of vehicle $i$ is chosen to be equal to the degree of vertex $i$. Therefore, for vehicle $i$ to satisfy its safety packet requirements, it must be scheduled at all time slots that are incident to it in the corresponding graph. This construction is clearly performed in polynomial-time.

    On the one hand, assume that MIS is solved with its given instance. That is, assume that an independent set $\mathscr{I}$ of maximum cardinality is found. Then, we can schedule each vehicle $i\in\mathscr{I}$ at the time slot incident to it. In doing so, we obtain a maximum number of successfully received packets since the cardinality of $\mathscr{I}$ is maximal and the size of the corresponding safety packet is equal to the degree of each vertex. Further, since $\mathscr{I}$ is an independent set, each time slot is used by at most one vehicle. Thus, VAMP$^\prime$ is solved. 
    
    On the other hand, assume that VAMP$^\prime$ is solved with its created instance. That is a solution to VAMP$^\prime$ in which a maximum number of successfully received safety packets is found. Then, since each scheduled vehicle respects the size of its safety packet, it must therefore be scheduled at all time slots incident to it. This set of scheduled vehicles forms an independent set since each time slot is used by at most one vehicle. Moreover, this independent set is of maximum cardinality since the number of successfully received safety packets is maximum.

    Putting all this together, we showed that MIS reduces to VAMP$^\prime$ in polynomial-time. This shows that VAMP$^\prime$ is NP-hard and thereby proves the lemma.
\end{IEEEproof}

In the sequel, we present the proposed solution to solve VAMP. First, we start by presenting a modified (adapted to specifically solve VAMP) version of the benchmark algorithm proposed in~\cite{7974737} to find the RBs scheduling. Next, we present our proposed DRL algorithm based on the \texttt{DDPG} method to find the transmission power and the association of the vehicle. 

Due to the broadcast nature of vehicular networks as well as for practical implementation reasons, the vehicle association optimization as formulated in~\eqref{mp1} is transformed into vehicle coverage optimization. In other words, vehicle $i$ communicates with vehicle $j$ at time slot $s$ if and only if the distance between the two vehicles at time slot $s$, $d_{i\rightarrow j}^{(s)}$, is at most $r_i$, i.e., $d_{i\rightarrow j}^{(s)}\leqslant r_i$. Here, $r_i$ is an optimization variable associated with vehicle $i$ and it denotes its communication range, i.e., vehicle $i$ can transmit to all vehicles $j\in\mathscr{N}_i$, where $\mathscr{N}_i$ is the set of neighbours of vehicle $i$ defined as follows: 
\begin{align}
	\label{n1}
	\mathscr{N}_i\coloneq\{j\in\mathscr{V}:d_{i\rightarrow j}^{(s)}\leqslant r_i\}\backslash\{i\}.
\end{align}
For example, in Fig.~\ref{sysmod1} the set $\mathscr{N}_W$ is given by the two red vehicles in the broadcasting range of $W$.

\section{The Deep Reinforcement Learning Solution}\label{sec:sol}
As discussed earlier, VAMP is NP-hard and is MINLP. Machine learning, in particular DRL, has received a lot of attention for solving NP-hard problems in wireless and vehicular networks. The flexibility in the design of the reward function and the action and state spaces is what makes DRL approaches attractive. 

To effectively solve VAMP, we propose a DRL-based approach using MDP. Note that the MDP  formulation could be helpful in exploring dynamic programming methods in future. Due also to the structure of the recursive relation found in the AoI function, exploring dynamic programming is appealing. The design of dynamic programming algorithms will be left for our future work and we only focus here on designing DRL solutions. First, we model VAMP as a single-agent MDP in which the RSU, as a DRL agent, interacts with the vehicular environment and makes decisions accordingly. The MDP is defined mathematically by the tuple $\mathcal{M}=(\mathcal{S},\mathcal{A},\mathcal{T},\mathcal{R})$, where $\mathcal{S}$ represents the set of states, $\mathcal{A}$ represents the set of actions, $\mathcal{T}$ represents the transition function, and $\mathcal{R}$ represents the reward function. In the following, we explicitly describe each element of MDP $\mathcal{M}$.
\begin{itemize}
	\item[$\mathcal{S}$:] a state $\mathbf{s}^{(t)}$ corresponds to a transmission period $t$. The DRL agent, once in-state $\mathbf{s}^{(t)}$, learns the status of the underlying vehicles at transmission period $t$. This includes vehicles' positions, directions, speeds, communication ranges, and transmission powers, the channel gains between vehicles, the current AoI of each pair of vehicles, and the current transmission period $t$. The terminal state $\mathbf{s}^{(n)}$ corresponds to the final transmission period $n$ including the status of the underlying vehicles.
	\item[$\mathcal{A}$:] an action is given by the matrix $\mathbf{A}=[\mathbf{a}_1,\mathbf{a}_2,\ldots,\mathbf{a}_m]^\top$ of size $m\times3$ where $\mathbf{a}_i=[a_{i1},a_{i2},a_{i3}]$ for $i\in\{1,2,\ldots,m\}$. The action $\mathbf{a}_i$ of vehicle $i\in\{1,2,\ldots,m\}$ is a row vector in which the variable $a_{i1}$ belongs to the real interval $[0, \overline{c}]$ that represents the communication coverage selected by the DRL agent for vehicle $i$, where $\overline{c}$ is an upper-bound for the communication range of each vehicle. Here, $a_{i1}=0$ means that the corresponding vehicle is selected to be a receiving vehicle in the corresponding transmission period. The variable $a_{i2}$ belongs to the set $\{0,1,\ldots,k\}$ and represents the time slots selected by the DRL agent for vehicle $i$. Finally, the variable $a_{i3}$ belongs to the real interval $[0, \overline{p}]$ which represents the power levels among which the DRL agent would select the transmission power for vehicle $i$. 
	\item[$\mathcal{T}$:] the transition from one state to another is deterministic. In other words, once the DRL agent is in state $\mathbf{s}^{(t)}$ and its chosen action is $\mathbf{A}^{(t)}$, it transitions to the next state $\mathbf{s}^{(t+1)}$ (to the next transmission period) with probability $1$ and observes the updates corresponding to the next state.
	\item[$\mathcal{R}$:] the reward is what the DRL agent seeks to maximize throughout its interaction with the environment. A straightforward approach to designing the reward is to choose the objective function in~\eqref{obj} since we seek to minimize the average AoI. Nevertheless, this reward function is not Markovian because the reward in a certain transmission period does not depend on the immediate state of the environment and on the chosen action. Rather, it depends on the history of the actions that the DRL agent has taken, i.e., the AoI between a pair of vehicles $(i,j)$, at transmission period $t$, clearly depends on whether or not the two vehicles have successfully delivered their safety packets in previous transmission periods. To cope with such subtlety, the reward function needs to be rethought. It is shown in~\cite{8904331} that maximizing the number of successfully delivered packets in each transmission period would help in minimizing the average AoI. Therefore, the reward function is chosen in proportion to the number of successfully delivered safety packets in each period: 
	\begin{align}
		\label{r1}
		\mathcal{R}(\mathbf{s}^{(t)},\mathbf{A}^{(t)})\coloneq r^{(t)}=\dfrac{\eta^{(t)}}{m^2},
	\end{align}
	where $\eta^{(t)}\coloneq2\sum_{i\in\mathscr{V}}\sum_{j\in\mathscr{V}}x_{i\rightarrow j}^{(t)}-m^2$ and the denominator $m^2$ is only used for normalization.
	
The rationale behind the choice of this reward function is that when all vehicles broadcast to each other (i.e., $x_{i\rightarrow j}^{(t)}=1$ for all $i$ and $j$), which is the preferred outcome, the reward will take on the maximum possible value which is $1$.  If the DRL agent maximizes the reward function in~\eqref{r1}, then it will maximize the number of successfully delivered safety packets. According to~\cite{8904331}, the DRL agent will therefore be able to reduce the average AoI across transmission periods.
\end{itemize}

The previously defined action space of MDP $\mathcal{M}$ consists of a mixture of discrete and continuous actions. Discrete action spaces make the learning procedure difficult due to the exponentially large number of possible actions. Exploring such a discrete action space will therefore be difficult for the DRL agent. In the case of VAMP, the number of discrete actions depends on the correspondence between the time slots $\{0,1,2,\ldots,k\}$ and the vehicles $\{1,2,\ldots,m\}$, which can be up to $\mathscr{O}(k^m)$. To overcome this issue, we employ a continuous control strategy for DRL, called deep deterministic policy gradient (\texttt{DDPG})~\cite{lillicrap2019continuous,9195789}.

Before describing the \texttt{DDPG} method in detail, we show how the time slots are matched to the vehicles and thus solving the discrete decisions making problem. To do so, we follow a benchmark algorithm proposed in~\cite{7974737}.

\subsection{Matching Vehicles to Time Slots}
For clarity, we describe the original algorithm proposed in~\cite{7974737} and include all necessary modifications and details. The main idea of the algorithm is to maximize the number of successfully received packets in each transmission period by applying a two-sided matching technique. We call this algorithm greedy matching (\texttt{GM}). If a vehicle is matched to a time slot, it is considered as a transmitting vehicle in that time slot, and otherwise, it is considered as a receiving vehicle in that time slot.

A two-sided matching $\Psi$ is defined mathematically as a mapping from $\mathscr{V}\cup\mathscr{S}\cup\varnothing$ to $\mathscr{V}\cup\mathscr{S}\cup\varnothing$, where $\Psi(i)\subseteq\mathscr{S}$, for $i\in\mathscr{V}$ and $\Psi(s)\subseteq\mathscr{V}$, for $s\in\mathscr{S}$. The vehicles that are matched to the same time slot are called \textit{matching peers}, which refer to an extension of the notion of \textit{roommates} in the original stable roommate matching problem~\cite{10.5555/68392}. A pair of vehicles $(i,i')$ is called a \textit{forbidden pair} if the distance between the two vehicles at time slot $s$ is upper-bounded by $r$, i.e., $d_{i\rightarrow i'}^{(s)}\leqslant r$. \texttt{GM} uses the notion of preference relation which we define in the following.

\subsection{Preference Relation}
Each vehicle establishes a preference relationship for all other vehicles. It must choose its corresponding matching peers in such a way that it is positioned away from them so that their communication ranges do not overlap. This reduces potential collision, which is defined in~\cite{7974737} as the cross-influence brought by multiple transmitting vehicles on the conflicting receiving vehicles. The cross-influence brought by any two (distinct) transmitting vehicles $i$ and $i'$ in time slot $s$ on the corresponding receiving vehicles is given as~\cite{7974737}:
\begin{align}
	\label{ci1}
	c_{i\rightarrow i'}^{(s)}\coloneq
	\begin{cases}
		(2r-d_{i\rightarrow i'}^{(s)})^2, & \text{if $2r>d_{i\rightarrow i'}^{(s)}$},\\
		\varepsilon, & \text{otherwise},
	\end{cases}
\end{align}
where $-0.1<\varepsilon<0$ is a sufficiently small constant. If a time slot is matched to only one vehicle, then the cross-influence exists and is equal to $\varepsilon$. Therefore, the average cross-influence brought by vehicle $i$ when matched to time slot $s$ is given by~\cite{7974737}:
\begin{align}
	\label{ci2}
	Q_i^{(s)}\coloneq\dfrac{1}{1+|\Psi(s)|}\sum_{i'\in\Psi(s)}c_{i\rightarrow i'}^{(s)}.
\end{align}
We can see from~\eqref{ci2} that, if $i$ is the only vehicle matched to time slot $s$, then the average cross-influence is equal to $\varepsilon$. More simply, the cross-influence is a way to measure how far two vehicles are. If they are far-way apart then the interference will be small and vice versa.

Based on the definition of the cross-influence, vehicle $i$ can construct its preference relationship. Specifically, vehicle $i$ prefers time slot $s$ to time slot $s'$ if (i) its average cross-influence that it brought at time slot $s$ is less than the one that it brought at time slot $s'$, i.e., $Q_i^{(s)}<Q_{i}^{(s')}$, (ii) there is no forbidden pair at time slot $s$, and (iii) time slot $s$ does not match more than $\overline{m}$ vehicles. Accordingly, the vehicles can define their feasible preference relations. \texttt{GM} goes through two phases: In the first phase, it finds a feasible solution, and in the second phase, it rotates the matching peers to further reduce their average cross-influence while guaranteeing feasibility.

\subsection{Phase 1: Finding a Feasible Solution}
A greedy approach is used in \texttt{GM} to find an initial solution~\cite{7974737}. Let $\Psi$ represent the current matching. If a not-yet-matched vehicle $i$ can form a forbidden pair with vehicle $i'\in\Psi(s)$, then vehicle $i$ cannot be matched to time slot $s$. We associate to each vehicle $i$ a set $\mathscr{M}_i$ of available matched time slots (the set of time slots that do not form forbidden pairs with $i$), defined as follows:
\begin{align}
	\label{Mj1}
	\mathscr{M}_i\coloneq\{s\in\mathscr{S}_{\text{matched}}:\Psi(s)\cap\mathscr{F}_i^{(s)}=\varnothing\},
\end{align}
where $\mathscr{S}_{\text{matched}}$ defines the set of time slots already-matched in the current matching $\Psi$ and $\mathscr{F}_i^{(s)}$ defines the set of vehicles that form forbidden pairs with vehicle $i$ at time slot $s$. In the case where $\mathscr{M}_i=\varnothing$, i.e., vehicle $i$ is not allowed to match to an already-matched time slot, it matches to a not-yet-matched time slot $s'$, that is, $\Psi(s')=\varnothing$. Otherwise, vehicle $i$ matches to time slot $s_i^*\in\mathscr{M}_i$ such that the average cross-influence brought by vehicle $i$ is as small as possible, i.e.,
\begin{align}
	\label{minq1}
	s^*_i\coloneq\text{arg}\min_s Q_i^{(s)}.
\end{align}
Once all vehicles or time slots are matched, the second phase begins.

\subsection{Phase 2: Rotating the Matching}
Consider a matching $\Psi$ with $L\geqslant2$ matched vehicles returned from the first phase. A single matched vehicle from $\mathscr{L}\coloneq\{1,\ldots,L\}$ is denoted by $\sigma(\ell)$ for $\ell\in\mathscr{L}$. We obtain the following original matching sequence:
\begin{align}
	\label{ms1}
	\bigl\{(\sigma(1),\Psi(\sigma(1))),\cdots,(\sigma(L),\Psi(\sigma(L)))\bigr\}.
\end{align}
An example of a matching for $L=3$ is given by $\bigl\{(1,\{1\}),(2,\{2\}),(3,\{3\})\bigr\}$.

A rotation sequence with respect to $\ell$, $\ell\in\mathscr{L}$, is defined as follows~\cite{7974737}:
\begin{align}
	\label{rs1}
	\mathscr{R}_\ell\coloneq\bigl\{(\sigma(i),\Psi(\sigma(\text{mod}_L(\ell+i)))):i\in\mathscr{L}\bigr\},
\end{align}
where the function $\text{mod}_L(x)$ denotes the modulo operator that returns the remainder of the division of $x$ by $L$. Note that $\mathscr{R}_L$ consists of the original matching sequence given in~\eqref{ms1}. For example, with the matching sequence $\bigl\{(1,\{1\}),(2,\{2\}),(3,\{3\})\bigr\}$, we obtain $\mathscr{R}_1=\bigl\{(1,\{2\}),(2,\{3\}),(3,\{1\})\bigr\}$, and $\mathscr{R}_2=\bigl\{(1,\{3\}),(2,\{1\}),(3,\{2\})\bigr\}$, and $\mathscr{R}_3=\bigl\{(1,\{1\}),(2,\{2\}),(3,\{3\})\bigr\}$.

Given a rotation sequence $\mathscr{R}_\ell$, $\ell\in\mathscr{L}\backslash\{L\}$, we define a rotation matching with respect to $\mathscr{R}_\ell$ as follows~\cite{7974737}:
\begin{align}
	\label{rm1}
	\Psi_{\mathscr{L},\mathscr{R}_\ell}\coloneq(\Psi\backslash\mathscr{R}_L)\cup\mathscr{R}_\ell.
\end{align}
The equation~\eqref{rm1} means that each vehicle $i$ is re-matched to the time slot defined in the rotation sequence $\mathscr{R}_\ell$, that is, vehicle $i$ is initially matched to time slot $\Psi(\sigma(i))$, and after the rotation matching, it will be matched to time slot $\Psi(\sigma(\text{mod}_L(\ell+i)))$. To illustrate this, let us take a look at the example below: $\Psi_{\{1,2,3\},\mathscr{R}_1}=(\Psi\backslash\mathscr{R}_3)\cup\mathscr{R}_1$. That is we transform $\Psi$ from the matching $\Psi(1)=\{1\},\Psi(2)=\{2\},\Psi(3)=\{3\}$ to the matching $\Psi_{\{1,2,3\},\mathscr{R}_1}(1)=\{2\},\Psi_{\{1,2,3\},\mathscr{R}_1}(2)=\{3\},\Psi_{\{1,2,3\},\mathscr{R}_1}(3)=\{1\}$ by removing $\mathscr{R}_3$ from $\Psi$ and adding $\mathscr{R}_1$ to it.

For $\ell\in\mathscr{L}\backslash\{L\}$, a rotation matching $\Psi_{\mathscr{L},\mathscr{R}_\ell}$ is valid if any vehicle $i\in\mathscr{L}$ does not form a forbidden pair with a matching peer in $\Psi_{\mathscr{L},\mathscr{R}_\ell}(\Psi_{\mathscr{L},\mathscr{R}_\ell}(i))\backslash\{i\}$. A valid rotation matching is optimal if it achieves the smallest average cross-influence, i.e., 
\begin{align}
	\label{om1}
	\ell^*=\text{arg}\min_{\ell\in\mathscr{L}}\sum\nolimits_{i\in\mathscr{L}}\sum\nolimits_{s\in\Psi_{\mathscr{L},\mathscr{R}_\ell}(i)}Q_i^{(s)}.
\end{align}
Finally, a matching $\Psi$ is called $L$-rotation stable if all rotation matchings in $\Psi$ are optimal, i.e., no rotation matching with $L\leqslant\overline{L}$ can further reduce the average cross-influence of $\Psi$, where $\overline{L}$ is used to control the complexity of the \texttt{GM} algorithm. 

With that said, the \texttt{GM} algorithm proceeds as follows. It is implemented in the RSU, which collects the position, distance, and speed information of the vehicles. Then it finds a feasible matching as described in the first phase. The second phase of \texttt{GM} contains several iterations. Each iteration consists of an optimal rotation matching that the RSU executes after selecting a random rotation sequence. The algorithm terminates when no rotation sequence further reduces the total cross-influence. \texttt{GM} is presented in pseudo-code in Algorithm~\ref{mgm1}.
\begin{algorithm}[ht!]
  \caption{The \texttt{GM} Algorithm}
  \label{mgm1}
  \begin{algorithmic}[1]
    \Require{Vehicles and time slots.}
    \Ensure{An $L$-rotation stable matching $\Psi$.}
    \LineComment{Initialization:}
    \State Let the current matching be $\Psi$.
    \State Define $\mathscr{S}'=\{1,2,\ldots,k\}$ as the not-yet-matched time slots.
    \For{$s=1,k$} 
    	\For{$i=1,m$}
	    	\State $\mathscr{F}_i^{(s)}=\{i'\in\mathscr{V}:d_{i\rightarrow i'}^{(s)}\leqslant r\}$.
		\EndFor
	\EndFor
    \LineComment{The first phase:}
    \For{$i=1,m$}
        \State Find the set of matched time slots $\mathscr{M}_i$ as in~\eqref{Mj1}.
        \If{$\mathscr{M}_i=\varnothing$ \textbf{and} $\mathscr{S}'\ne\varnothing$}
			\State Select a random time slot $s^*_i$ from $\mathscr{S}'$.
		\Else
			\State Select time slot $s^*_i$ according to~\eqref{minq1}.	
        \EndIf
        \State $\Psi(i)=s^*_i$.
        \State $\mathscr{S}'=\mathscr{S}'\backslash\{s^*_i\}$.
    \EndFor
    \LineComment{The second phase:}
    \While{\text{the average cross influence can be reduced}}
    	\State Select a random rotation sequence $\mathscr{R}_\ell$.
		\State Find the optimal rotation matching $\Psi^*_{\mathscr{L},\mathscr{R}_\ell}$.
		\State $\Psi=\Psi_{\mathscr{L},\mathscr{R}_\ell}^*$.
	\EndWhile 
  \end{algorithmic}
\end{algorithm}

The complexity of \texttt{GM} algorithm is upper-bounded by the complexity of the second phase in line 18. The complexity of the second phase is given by $\mathscr{O}(m^{\overline{L}})$ where $\overline{L}$ is the maximum length of a rotation sequence. In our vehicular network scenario, the value of $\overline{L}$ is fixed and is much smaller than $m$, thus the complexity $\mathscr{O}(m^{\overline{L}})$ is considered polynomial. The overall complexity of \texttt{GM} is thus given by $\mathscr{O}(\overline{T}m^{\overline{L}})$, where $\overline{T}$ is the maximum number of iterations needed for the \textbf{while} loop in line 18 of Algorithm~\ref{mgm1} to terminate. The space complexity of \texttt{GM} is $\mathscr{O}(mk)$ since we need to store the matching $\Psi$ and the set of matched slots as given in~\eqref{Mj1}.

To solve VAMP, the RSU (the DRL agent) applies the \texttt{DDPG} method (will be described shortly) in conjunction with \texttt{GM}. Since \texttt{GM} performs time slots/vehicles matching, thus the action space of MDP $\mathcal{M}$ defined earlier is modified to contain only coverage selection decisions and power allocation decisions. Note that, as we saw earlier, vehicle $i$ chooses an action $\mathbf{a}_i=[a_{i1},a_{i2},a_{i3}]$ for $i\in\{1,2,\ldots,m\}$. We modify this action by encoding it as $[a_{i1},a_{i3}]$ since the action $a_{i2}$ corresponds to the time slots/vehicles matching already chosen by \texttt{GM}. 

\begin{figure*}
	\centering
	\includegraphics[width=0.7\textwidth]{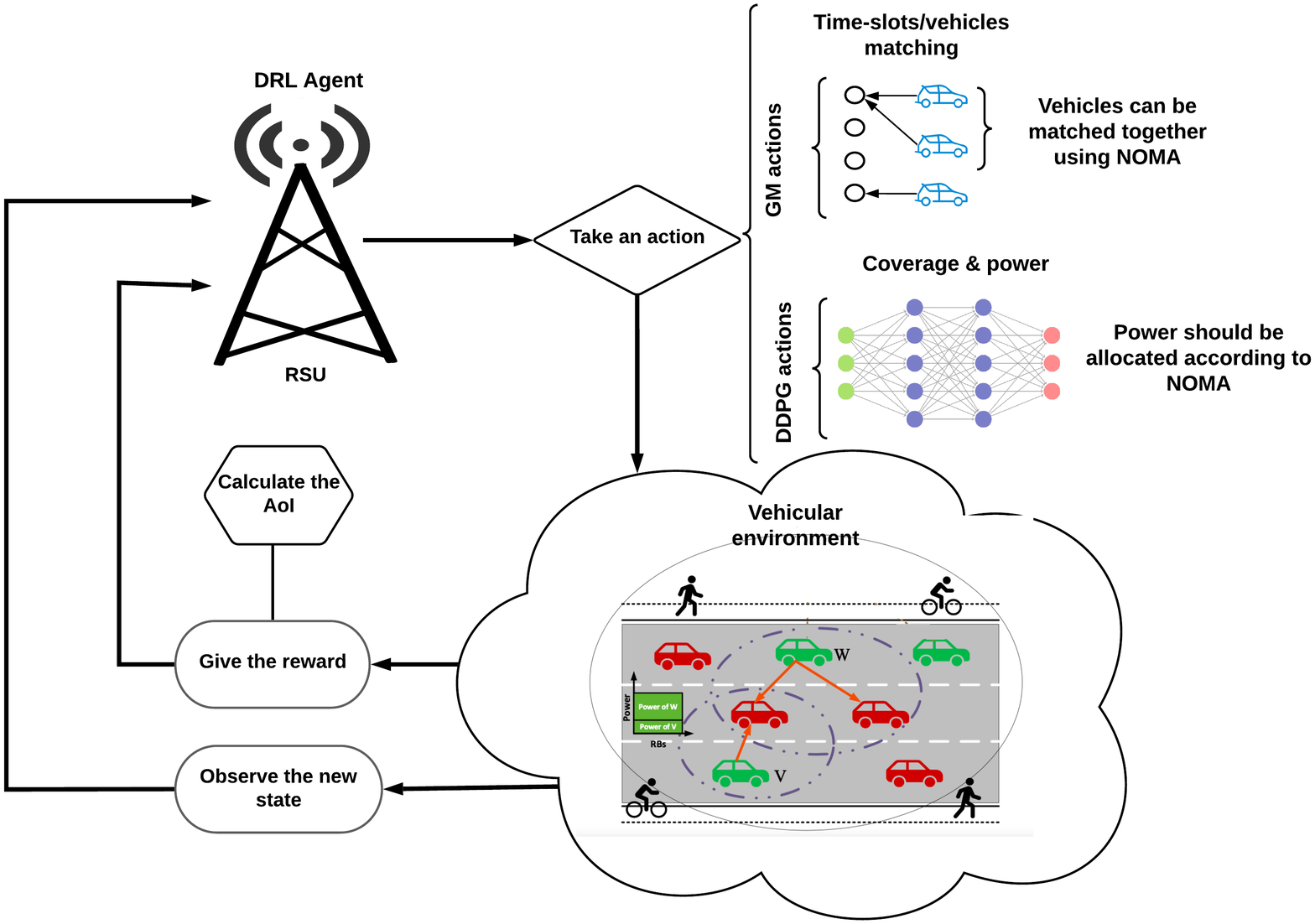}
	\caption{The interaction between the DRL agent and the vehicular environment. The DRL agent executes the \texttt{DDPG} method and the \texttt{GM} matching algorithm to select actions. Precisely, the actions of \texttt{GM} consist of allocating time slots to vehicles such that two or more vehicles can be grouped using NOMA while the actions of \texttt{DDPG} consist of allocating coverage and transmission power. The power should be allocated carefully to take advantage of NOMA. Once the action is taken, the reward is calculated to reflect the current AoI in the vehicular network.}
	\label{fig:sysblock}
\end{figure*}

The interaction between the DRL agent and the vehicular network environment is given in the system block in Fig.~\ref{fig:sysblock}.

The proposed \texttt{DDPG}-based DRL algorithm is called \texttt{DDPG-GM} and is implemented centrally in the RSU. It proceeds in $E$ episodes as follows. For each episode, the DRL agent observes the vehicular environment and makes decisions accordingly. The decision making is divided into two parts: \texttt{GM} actions (i.e., action $a_{i2}$ for each vehicle $i$) and \texttt{DDPG} actions (i.e., actions $[a_{i1}, a_{i3}]$ for each vehicle $i$). For \texttt{GM} actions, once the DRL agent observes the current state $\mathbf{s}^{(t)}$, it applies the \texttt{GM} matching algorithm to find the correspondence between time slots and vehicles. Note that, we can encode the resulting time slots/vehicles matching in the current state, i.e., the current state includes not only the elements of the state space $\mathcal{S}$ but also the resulting time slots/vehicles matching. Then, for \texttt{DDPG} actions, we apply an actor-critic, off-policy, model-free algorithm that operates on continuous action spaces~\cite{lillicrap2019continuous}. \texttt{DDPG} combines the advantages of DQN (including off-policy learning, replay buffer, and target Q-network updates) and the policy gradient to efficiently make continuous decisions. The DRL agent, to perform \texttt{DDPG} actions as shown in Fig.~\ref{fig:sysblock}, creates two deep neural networks (DNNs), one is called the actor-network and the other the critic network. The role of the actor-network is to learn the policy of the DRL agent by making decisions about the broadcast communication coverage of each vehicle as well as the transmission power of each transmitting vehicle. On the other hand, the role of the critic network is to learn the value of the Q-function for each state/action pair. Once the \texttt{DDPG-GM} algorithm has observed the environment and chosen its actions $\mathbf{A}^{(t)}$, the reward function $\mathcal{R}(\mathbf{s}^{(t)},\mathbf{A}^{(t)})$ is evaluated to find the number of successfully delivered safety packets. Then the environment transitions to the next state and the new time slots/vehicles matching is obtained, which is encoded in the new state $\mathbf{s}^{(t+1)}$. A replay buffer is then used to record the transition tuple $(\mathbf{s}^{(t)}, \mathbf{A}^{(t)}, \mathbf{s}^{(t+1)}, r^{(t)})$. Once the replay buffer contains a sufficient number of tuples, a mini-batch of a certain size is sampled from it to perform the policy gradient. Finally, the actor and critic parameters are updated by minimizing some loss functions. The \texttt{DDPG-GM} algorithm is illustrated in Algorithm~\ref{alg:mgm-ddpg} and some details about it will follow. 
\begin{algorithm}[ht!]
  \caption{The \texttt{DDPG-GM} Algorithm}
  \label{alg:mgm-ddpg}
  \begin{algorithmic}[1]
    \Require{The vehicular environment.}
    \Ensure{A learned policy for the time slots/vehicles matching, coverage areas, and transmission powers.}
    \LineComment{Initialization:}
    \State Initialize the critic network $Q(\mathbf{s},\mathbf{A}|\theta^Q)$.
    \State Initialize the actor network $\pi(\mathbf{s}|\theta^\pi)$.
    \State Initialize the critic target network $Q'$.
    \State Initialize the actor target network $\pi'$.
    \State Initialize the replay buffer.
    \State Set $\mathbf{r}_{\text{avg}}\gets[0,0,\ldots,0]$.
    \LineComment{Learning:}
    \For{$e=1,E$}
        \State Reset the vehicular environment.
        \State Apply the \texttt{GM} algorithm.
        \State Observe the first state $\mathbf{s}^{(1)}$.
        \State Initialize a random process for \texttt{DDPG} exploration.
        \State Set $r_e\gets0$.
        \For{$t=1,n-1$}
			\State Choose a \texttt{DDPG} action $\mathbf{A}^{(t)}$ by called $\pi(\mathbf{s}^{(t)}|\theta^\pi)$.
			\State Perform a preprocessing for $\mathbf{A}^{(t)}$.
			\State Find the reward $r^{(t)}$.
			\State Accumulate the reward $r_e\gets r_e+r^{(t)}$.
			\State Apply the \texttt{GM} algorithm.
	        \State Observe the next state $\mathbf{s}^{(t+1)}$.
	        \State Record $(\mathbf{s}^{(t)}, \mathbf{A}^{(t)}, \mathbf{s}^{(t+1)}, r^{(t)})$ in the replay buffer.
	        \State Sample $T$ transitions from the replay buffer.
	        \State Set $v^{(t)}\gets r^{(t)} + \gamma Q(\mathbf{s}^{(t+1)}, \pi'(\mathbf{s}^{(t+1)}|\theta^{\pi'})|\theta^{Q'})$.
	        \State Update $\theta^Q$ by minimizing the following loss: \[\frac{1}{T}\sum_{t}(v^{(t)}-Q(\mathbf{s}^{(t)},\mathbf{A}^{(t)}|\theta^Q))^2.\]
	        \State Update $\theta^\pi$ using the policy gradient method: \[\frac{1}{T}\sum_t\nabla_{\mathbf{A}}Q(\mathbf{s},\mathbf{A}|\theta^Q)|_{\mathbf{s}=\mathbf{s}^{(t)},\mathbf{A}=\pi(\mathbf{s}^{(t)})}\nabla_{\theta^{\pi}}\pi(\mathbf{s}|\theta^\pi)|_{\mathbf{s}^{(t)}}.\]
	        \State Update the target networks weights: 
	        \begin{align*}
	        	&\theta^{Q'}\gets\iota\theta^Q+(1-\iota)\theta^{Q'}.\\
				&\theta^{\pi'}\gets\iota\theta^\pi+(1-\iota)\theta^{\pi'}.
	        \end{align*}    
	        \State Set $\mathbf{s}^{(t)}\gets\mathbf{s}^{(t+1)}$.
        \EndFor
        \State Set $\mathbf{r}_{\text{avg}}[e]\gets r_e$.
        \If{$e\geqslant50$}
        	\State Return the average rewards $\text{mean}(\mathbf{r}_{\text{avg}}[e-49:e])$.
        \EndIf
    \EndFor
  \end{algorithmic}
\end{algorithm}

The \texttt{DDPG-GM} algorithm is divided into two main phases: an initialization phase and a learning phase. During the initialization phase, the creation of DNNs is performed. First, a critic network $Q(\mathbf{s},\mathbf{A}|\theta^Q)$ with weights $\theta^Q$ is created for each state $\mathbf{s}$ and action $\mathbf{A}$. Then, an actor-network $\pi(\mathbf{s}|\theta^\pi)$ with weights $\theta^\pi$ is created for each state $\mathbf{s}$. Two target networks are also created that correspond to the critic and actor networks. The target critic network is denoted as $Q'$ and has weights $\theta^{Q'}$, which are initially set to $\theta^Q$ and the target actor-network is denoted as $\pi'$ and has weights $\theta^{\pi'}$, which are initially set to $\theta^\pi$. Finally, the replay buffer is initialized to contain empty transitions, and a $E\times1$ reward vector $\mathbf{r}_{\text{avg}}$ is initialized to zero to contain the average rewards of the last 50 episodes. 

During the learning phase, \texttt{DDPG-GM} iterates over $E$ episodes. For each episode $e$, the vehicular network is reset, including the positions, speeds, etc. of the vehicles. Then the \texttt{GM} algorithm is applied to obtain the time slots/vehicles matching solution. This solution is encoded with the other parameters of the vehicles to obtain the current state of the environment. Then a random process is initialized to perform the action exploration in \texttt{DDPG} and the reward $r_e$ that will take into account the reward accumulated so far is initialized to zero. From this point, \texttt{DDPG-GM} iterates through the transmission periods until the second-to-last transmission period and it performs action exploration and policy gradient. An action $\mathbf{A}^{(t)}$ is selected using the actor-network as follows:
\begin{align}\label{actor1}
	\mathbf{A}^{(t)}=\pi(\mathbf{s}^{(t)}|\theta^\pi),
\end{align}
where $t$ denotes the transmission period. The action is then pre-processed to obtain a feasible action to VAMP and the reward function is invoked to compute the number of successfully delivered safety packets as in~\eqref{r1}. Then the accumulated reward $r_e$ is updated and the next state $\mathbf{s}^{(t+1)}$ is obtained by first applying the \texttt{GM} algorithm to find the corresponding time slots/vehicles matching during $t+1$, and then observing the vehicular environment. Once the transition tuple $(\mathbf{s}^{(t)}, \mathbf{A}^{(t)}, r^{(t)}, \mathbf{s}^{(t+1)})$ is found, it is stored in the replay buffer. As soon as the replay buffer is not empty, a mini-batch training step is performed by sampling a tuple of random $T$ transitions from the replay buffer. This mini-batch of transitions is trained by minimizing the minimum square error of the loss function:
\begin{align}\label{loss1}
	\frac{1}{T}\sum_{t}(v^{(t)}-Q(\mathbf{s}^{(t)},\mathbf{A}^{(t)}|\theta^Q))^2,
\end{align}
where $v^{(t)}=r^{(t)} + \gamma Q(\mathbf{s}^{(t+1)}, \pi'(\mathbf{s}^{(t+1)}|\theta^{\pi'})|\theta^{Q'})$ and $\gamma$ denotes the discount value. The minimization performed in~\eqref{loss1} updates the weights $\theta^Q$ of the critic network. Then the weights $\theta^\pi$ of the actor network are updated by performing the gradient descent method as follows:
\begin{align}\label{gd1}
	\frac{1}{T}\sum_t\nabla_{\mathbf{A}}Q(\mathbf{s},\mathbf{A}|\theta^Q)|_{\mathbf{s}=\mathbf{s}^{(t)},\mathbf{A}=\pi(\mathbf{s}^{(t)})}\nabla_{\theta^{\pi}}\pi(\mathbf{s}|\theta^\pi)|_{\mathbf{s}^{(t)}}.
\end{align}
Finally, the target network weights $\theta^{Q'}$ and $\theta^{\pi'}$ are updated using the hyperparameter $\iota$, the previous state is updated to contain the next state, and the average reward of the last 50 episodes is computed.

\section{Simulation Results}\label{sec:sim}
In this section, we validate the proposed \texttt{DDPG}-based resource allocation method in a simulated vehicular network. The simulation setup is based on the highway case detailed in 3GPP TR 37.885~\cite{3gpp.37.885}. We used the Monte Carlo simulation method to perform all the simulations. We used the Julia programming language~\cite{bezanson2017julia} with the Flux.jl~\cite{innes:2018} machine learning library package to perform deep learning. The training is conducted in an offline fashion but the validation is done in an online fashion where the vehicles and network parameters change in real-time. We used a desktop computer with 16 GB of RAM and with a 2,6 GHz Intel i7 processor and NVIDIA GeForce RTX 2070 Super graphic card.

We consider a multi-lane highway with a total length of $2$ km where each lane has a width $4$ meters. There are a total of six lanes---three for the forward direction (vehicles move from right to left) and three for the backward direction (vehicles move from left to right). The vehicles are generated in the vehicular environment according to a spatial Poisson process. The vehicle speed determines the vehicle density, and the average distance between vehicles (in the same lane) is set to $2.5\text{s}\times v$~\cite{7974737} where $v$ is the absolute vehicle speed. The speed of a vehicle depends on the lane it is in: the $i$th forward lane (top to bottom with $i\in\{1,2,3\}$) is characterized by the speed of $60+2(i-1)\times10$ km/h, while the $i$th backward lane (top to bottom with $i\in\{1,2,3\}$) is characterized by the speed of $100-2(i-1)\times10$ km/h. Unless otherwise specified, the important simulation parameters are given in table~\ref{my-label}.
\begin{table}[htp!] 
 \caption{Vehicular network parameters}
\label{my-label}
\centering
\resizebox{1\columnwidth}{!}{%
\begin{tabular}{||l||l||}
\hline
Parameter                                 & Value \\ [0.5ex]
\hline\hline       
Carrier frequency                         & $2$ GHz \\ \hline
Bandwidth per RB                          & $100$ kHz \\ \hline
Vehicle antenna height                    & $1.5$ m \\ \hline
Vehicle antenna gain                      & $3$ dBi \\ \hline
Vehicle receiver noise figure             & $9$ dB \\ \hline
Shadowing distribution                    & Log-normal \\\hline
Fast fading                               & Rayleigh fading \\\hline
Pathloss model                            & LOS in WINNER + B1~\cite{winner} \\\hline
Shadowing standard deviation              & $3$ dB \\ \hline
Road configuration                        & Highway road configuration~\cite{3gpp.37.885} \\ \hline
Vehicle drop model                        & Spatial Poisson process \\\hline
Number of transmission periods $n$        & $200$ \\\hline
Safety packets sizes $\zeta$              & $[3,15]$ KB \\\hline
Maximum coverage $\overline{c}$           & $200$ m\\\hline
Maximum transmission power $\overline{p}$ & $1$ Watt \\\hline
Noise power $N_0$                         & $-114$ dBm \\ [1ex]
\hline
\end{tabular}
}
\end{table}

\begin{table*}
 \caption{Actor and critic parameters}
\label{tab:actor}
\centering
\begin{tabular}{||l||l||}
\hline
Actor Parameter                               & Value \\ [0.5ex]
\hline\hline       
Size of the input layer                 & $210$ or $1062$\\\hline
Size of the output layer                & $8$ or $20$\\\hline
Number of hidden layers (HLs)           & $2$ \\ \hline
Number of neurons of first HL           & $500$ \\ \hline
Number of neurons of second HL          & $300$ \\ \hline
Activation function of the first HL     & relu \\ \hline
Activation function of the second HL    & relu \\ \hline
Activation function of the output layer & tanh \\ \hline
Target update factor $\iota$            & 1\text{e}-3 \\\hline
Learning rate                           & 2\text{e}-4 \\ [1ex]
\hline
\end{tabular}
\quad
\begin{tabular}{||l||l||}
\hline
Critic Parameter                               & Value \\ [0.5ex]
\hline\hline       
Size of the input layer                 & $218$ or $1082$\\\hline
Size of the output layer                & $1$\\\hline
Number of HLs                           & $2$ \\ \hline
Number of neurons of the first HL       & $500$ \\ \hline
Number of neurons of the second HL      & $300$ \\ \hline
Activation function of the first HL     & relu \\ \hline
Activation function of the second HL    & relu \\ \hline
Activation function of the output layer & nothing \\ \hline
Target update factor $\iota$            & 1\text{e}-3 \\\hline
Learning rate                           & 1\text{e}-4 \\ [1ex]
\hline
\end{tabular}

\end{table*}

We train two different vehicular networks. The first one is composed of $m=4$ vehicles (called a $4$-sized vehicular network) while the second network is composed of $m=10$ vehicles (called a $10$-sized vehicular network). The DNNs are created and trained in the Julia programming language~\cite{bezanson2017julia} using Flux.jl~\cite{innes:2018} machine learning library. The actor DNN consists of an input layer of size $\sigma_{\text{actor}}=2m^2k + m^2 + 3mk + 4m + 2$ and an output layer of size $\alpha_{\text{actor}}=2m$ and two fully connected hidden layers containing $500$ and $300$ neurons respectively. The rectified linear unit activation function (relu) given by $\max\{0,x\}$ is used in each layer except in the last one in which the hyperbolic tangent function (tanh) given by $(\exp(2x)-1)/(\exp(2x)+1)$ is used. The critic DNN consists of an input layer of size $\sigma_{\text{actor}}+\alpha_{\text{actor}}$ and an output layer of size 1 and two fully connected hidden layers containing $500$ and $300$ neurons respectively. The relu activation function is used in each layer except in the last one where no activation function is used. The actor and critic DNNs are trained with the ADAM optimizer~\cite{adam} with a corresponding learning rate of 2\text{e}-4 and 1\text{e}-4. The training lasts $E=1000$ episodes. The target update hyperparameter $\iota$ is set equal to $\iota =$ 1\text{e}-3 and the discount factor $\gamma$ is set to $\gamma=$ 99\text{e}-2. The replay buffer has size of 3\text{e}6 and the batch size is $64$. The random noise exploration parameter is set to 1\text{e}-1. The channel coefficients are time-varying in each time slot of length $25$ ms and a transmission period has $100$ ms length.

We summarize the parameters of the actor and critic DNNs in table~\ref{tab:actor}, where the size of the input layer depends on the trained vehicular network (either a $4$-sized vehicular network or a $10$-sized vehicular network).


In the literature, we mentioned that very few papers study the AoI optimization problem in a vehicular network. Further, no previous work studied the objective of minimizing the average AoI by optimizing vehicles coverage, resource allocation (RBs and power), and half-duplex transmitter selection. Our work can be seen as an extension of~\cite{7974737} where the authors proposed how to solve the half-duplex transmitter selection problem but without optimizing the AoI. We adapted the proposed algorithm in~\cite{7974737} and build on it to design a DRL method that solves a challenging joint discrete/continuous optimization problem to minimize the AoI. In the simulations results, we compared our proposed solutions to three baseline approaches that we adapted and implemented to solve VAMP: two are based on the NOMA technique and one is based on the OMA technique. The main idea of all baselines comes from~\cite{7974737} and they solve the time slots/matching problem using the \texttt{GM} algorithm. All baselines randomly select the coverage areas of the different vehicles in the interval $[0,\overline{c}]$. All baselines are centralized and implemented inside the RSU. They are called \texttt{MAX-GM}, \texttt{RND-GM}, and \texttt{OMA-GM}. In the \texttt{OMA-GM}, every time slot is used by at most one vehicle to meet OMA constraints and the vehicles transmit with their maximum transmission power $\overline{p}$. In the \texttt{MAX-GM} and \texttt{RND-GM}, every time slot can be used by any vehicle under the NOMA technique. However, in the \texttt{MAX-GM}, vehicles transmit with their maximum transmission power $\overline{p}$, while in the \texttt{RND-GM}, vehicles transmit with random transmission powers in the interval $[0,\overline{p}]$.

\begin{figure}[ht!]
  \centering
  \captionsetup{justification=centering,margin=1cm}
  \resizebox{.4\textwidth}{!}{%
    \begin{tikzpicture}
    \begin{axis}[
        xlabel={Episodes ($\times10$)},
        ylabel={Avg. rewards of the last $50$ episodes},
        set layers,
        grid=both,
        ymin=-172.1,ymax=-168.4,
        xticklabels={5,15,25,35,45,55,65,75,85,95},
        xtick={50,150,250,350,450,550,650,750,850,950},
        xmin=50,xmax=950,
        x label style={font=\footnotesize},
        y label style={font=\footnotesize}, 
        ticklabel style={font=\footnotesize},
        legend style={at={(0.301,1)},font=\scriptsize},]
    \addplot[blue,dashed,line width=1pt] table [x=epochs, y=AoI, col sep=comma,] {RND_Reward_10.csv};
    \addlegendentry{\texttt{DDPG-GM}};
    \end{axis}
    \end{tikzpicture}
  }
  \caption{The training rewards averaged over the last 50 episodes for the $10$-sized network when varying the number of episodes.}
  \label{fig:1}
\end{figure}
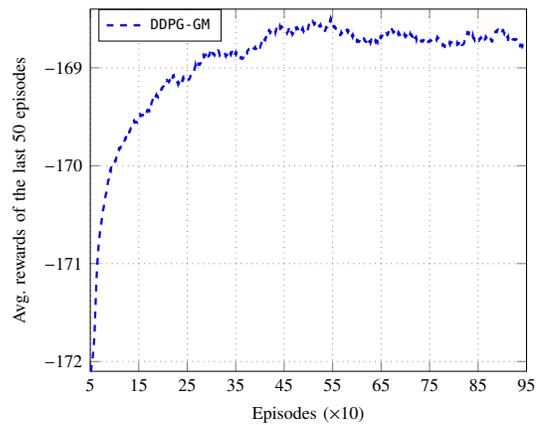

From Fig.~\ref{fig:1}, we conclude that:
\begin{itemize}
    \item The rewards increase as the number of episodes increases, which illustrate the convergence of the proposed \texttt{DDPG-GM} algorithm;
    \item The accumulated rewards do not vary a lot despite the highly dynamic nature of the network. This illustrates the stability of our solutions.
\end{itemize}


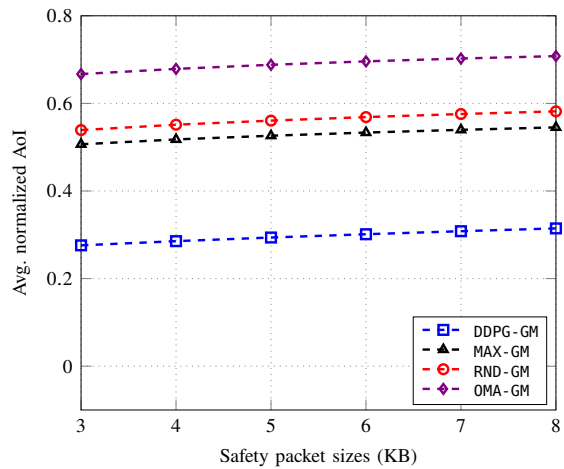
\begin{figure}[h!]
  \centering
  \captionsetup{justification=centering,margin=1cm}
  \resizebox{.42\textwidth}{!}{%
    \begin{tikzpicture}[spy using outlines={circle,magnification=8,size=2cm,connect spies}]
    	\begin{axis}[
		axis on top,
        xlabel={Safety packet sizes (KB)},
        ylabel={Avg. normalized AoI},
        grid=both,
        ymax=0.8,ymin=-0.1,
        xticklabels={3,4,5,6,7,8},
        xtick={300,400,500,600,700,800},
        xmin=300,xmax=800,
        x label style={font=\footnotesize},
        y label style={font=\footnotesize}, 
        ticklabel style={font=\footnotesize},
        legend style={at={(1,1)},font=\scriptsize},]
    \addplot[blue,mark=square,line width=1pt,dashed,mark options=solid] coordinates {
		(300.0, 0.2759271) (400.0, 0.28545812) (500.0, 0.2937606) (600.0, 0.3012674) (700.0, 0.3081858) (800.0, 0.3145791)
    };\label{ddpg10c200};
    \addplot[black,mark=triangle,line width=1pt,dashed,mark options=solid] coordinates {
		(300, 0.50661516) (400, 0.5176976) (500, 0.5262076) (600, 0.5333665) (700, 0.53961533) (800, 0.5452263)
    };\label{maxmgm10c200};
    \addplot[red,mark=o,line width=1pt,dashed,mark options=solid] coordinates {
		(300, 0.5391069) (400, 0.5513418) (500, 0.5606522) (600, 0.56871855) (700, 0.5757197) (800, 0.58183646)
    };\label{mgm10c200};
    \addplot[violet,mark=diamond,line width=1pt,dashed,mark options=solid] coordinates {
		(300.0, 0.6669472) (400.0, 0.6787875) (500.0, 0.6881309) (600.0, 0.69593066) (700.0, 0.7024545) (800.0, 0.70795566)
    };\label{mgmoma10c200};
	\end{axis}
    \node [draw,fill=white] at (rel axis cs: 0.84,0.125) {\shortstack[l]{
    \ref{ddpg10c200} \scriptsize\texttt{DDPG-GM}\\
    \ref{maxmgm10c200} \scriptsize\texttt{MAX-GM} \\
    \ref{mgm10c200} \scriptsize\texttt{RND-GM}\\
    \ref{mgmoma10c200} \scriptsize\texttt{OMA-GM}}};
    \end{tikzpicture}
  }
  \caption{The average normalized AoI vs. the safety packet sizes for the $10$-sized network.}
  \label{fig:1b}
  \end{figure}%

	\begin{figure}[h!]
	  \centering
  \captionsetup{justification=centering,margin=1cm}
  \resizebox{.42\textwidth}{!}{%
    \begin{tikzpicture}[spy using outlines={circle,magnification=8, connect spies}]
    	\begin{axis}[
        xlabel={Safety packet sizes (KB)},
        ylabel={Avg. normalized AoI},
        grid=both,
        ymax=0.6,ymin=-0.1,
        xticklabels={3,4,5,6,7,8},
        xtick={300,400,500,600,700,800},
        xmin=300,xmax=800,
        x label style={font=\footnotesize},
        y label style={font=\footnotesize}, 
        ticklabel style={font=\footnotesize},
        legend style={at={(1,1)},font=\scriptsize},]
    \addplot[blue,mark=square,line width=1pt,dashed,mark options=solid] coordinates {
		(300.0, 0.2969804) (400.0, 0.30246457) (500.0, 0.30693126) (600.0, 0.31105042) (700.0, 0.31457126) (800.0, 0.31745666)
    };\label{ddpg4c200};
    \addplot[black,mark=triangle,line width=1pt,dashed,mark options=solid] coordinates {
		(300.0, 0.4317108) (400.0, 0.44464082) (500.0, 0.45384166) (600.0, 0.46235874) (700.0, 0.4688342) (800.0, 0.4751196)
    };\label{maxmgm4c200};
    \addplot[red,mark=o,line width=1pt,dashed,mark options=solid] coordinates {
		(300.0, 0.47588998) (400.0, 0.4890275) (500.0, 0.5007562) (600.0, 0.5105454) (700.0, 0.5181317) (800.0, 0.525325)
    };\label{mgm4c200};
    \addplot[violet,mark=diamond,line width=1pt,dashed,mark options=solid] coordinates {
		(300.0, 0.51989543) (400.0, 0.53228956) (500.0, 0.5414425) (600.0, 0.5495033) (700.0, 0.5560213) (800.0, 0.5625359)
    };\label{mgmoma4c200};
    \end{axis}
    \node [draw,fill=white] at (rel axis cs: 0.84,0.125) {\shortstack[l]{
    \ref{ddpg4c200} \scriptsize\texttt{DDPG-GM}\\
    \ref{maxmgm4c200} \scriptsize\texttt{MAX-GM}\\
    \ref{mgm4c200} \scriptsize\texttt{RND-GM}\\
    \ref{mgmoma4c200} \scriptsize\texttt{OMA-GM}}};
    \end{tikzpicture}
  }
  \caption{The average normalized AoI vs. the safety packet sizes for the $4$-sized network.}
  \label{fig:1a}
 	\end{figure}
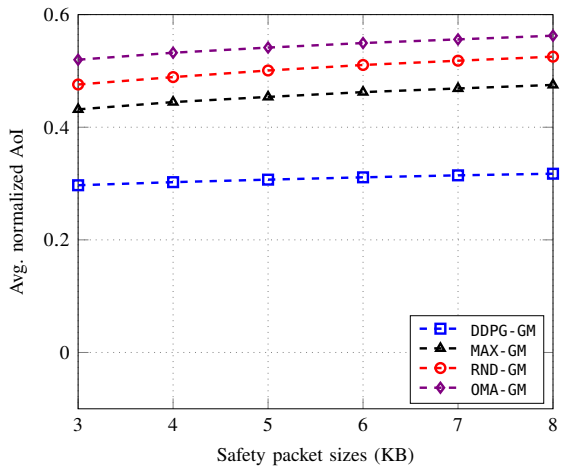

In Fig.~\ref{fig:1b} and Fig.~\ref{fig:1a}, we illustrate the performance of \texttt{DDPG-GM} when varying the size of the vehicular network as well as the size of the safety packets. We conclude:
\begin{itemize}
    \item As the size of the safety packets increases, the AoI increases (although the increase is only marginal) as it becomes more difficult to satisfy their requirements. Thus, fewer safety packets will be successfully broadcast to their destinations, and thus the time between the generation of safety packets and their successful delivery will become large, which increases AoI;
    \item \texttt{DDPG-GM} manages to learn a better AoI compared to other algorithms and is, therefore, able to deliver more safety packets by optimizing the coverage areas and transmission powers of each vehicle;
    \item The AoI is higher when the size of the vehicle network is large due to the half-duplex and broadcast nature of V2X communications. Indeed, when there are more vehicles in the network, if any of them broadcast their safety packets, then the interference degrades the performance and thus fewer safety packets will be successfully delivered;
    \item The OMA technique has the largest AoI, due to the inefficient use of time slots resources.
\end{itemize}

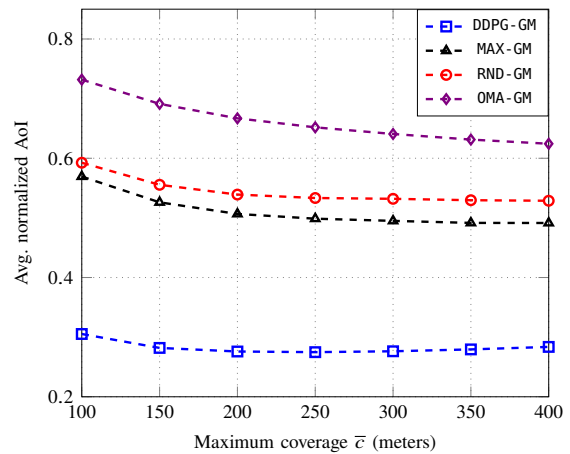
\begin{figure}[h!]
  \centering
	  \captionsetup{justification=centering,margin=1cm}
  \resizebox{.42\textwidth}{!}{%
    \begin{tikzpicture}
    	\begin{axis}[
		xlabel={Maximum coverage $\overline{c}$ (meters)},
        ylabel={Avg. normalized AoI},
        grid=both,
        ymax=0.85,ymin=0.2,
        xticklabels={100,150,200,250,300,350,400},
        xtick={100,150,200,250,300,350,400},
        xmin=100,xmax=400,
        x label style={font=\footnotesize},
        y label style={font=\footnotesize}, 
        ticklabel style={font=\footnotesize},
        legend style={at={(1,1)},font=\scriptsize},]
    \addplot[blue,mark=square,line width=1pt,dashed,mark options=solid] coordinates {
		(100.0, 0.30529246) (150.0, 0.28194815) (200.0, 0.2759271) (250.0, 0.27484006) (300.0, 0.27632067) (350.0, 0.2794552) (400.0, 0.28364184)
    };\addlegendentry{\texttt{DDPG-GM}};
    \addplot[black,mark=triangle,line width=1pt,dashed,mark options=solid] coordinates {
		(100.0, 0.569335) (150.0, 0.52608335) (200.0, 0.50661516) (250.0, 0.49865675) (300.0, 0.49498394) (350.0, 0.49158964) (400.0, 0.4914435)
    };\addlegendentry{\texttt{MAX-GM}};
    \addplot[red,mark=o,line width=1pt,dashed,mark options=solid] coordinates {
		(100.0, 0.5924982) (150.0, 0.5554778) (200.0, 0.5391069) (250.0, 0.5333448) (300.0, 0.5319808) (350.0, 0.52964187) (400.0, 0.5287235)
    };\addlegendentry{\texttt{RND-GM}};
    \addplot[violet,mark=diamond,line width=1pt,dashed,mark options=solid] coordinates {
		(100.0, 0.73182434) (150.0, 0.69119865) (200.0, 0.6669472) (250.0, 0.6518346) (300.0, 0.64070904) (350.0, 0.6313036) (400.0, 0.6243216)
    };\addlegendentry{\texttt{OMA-GM}};
    \end{axis}
    \end{tikzpicture}
  }
  \caption{The average normalized AoI vs. the maximum coverage distance for the $10$-sized network.}
  \label{fig:2a}
 \end{figure}

\begin{figure}[h!]
  \centering
  \captionsetup{justification=centering,margin=1cm}
  \resizebox{.42\textwidth}{!}{%
    \begin{tikzpicture}
    	\begin{axis}[
	    xlabel={Maximum coverage $\overline{c}$ (meters)},
        ylabel={Avg. normalized AoI},
        grid=both,
        ymax=0.8,ymin=0.2,
        xticklabels={100,150,200,250,300,350,400},
        xtick={100,150,200,250,300,350,400},
        xmin=100,xmax=400,
        x label style={font=\footnotesize},
        y label style={font=\footnotesize}, 
        ticklabel style={font=\footnotesize},
        legend style={at={(1,1)},font=\scriptsize},]
    \addplot[blue,mark=square,line width=1pt,dashed,mark options=solid] coordinates {
		(100.0, 0.30973583) (150.0, 0.29984206) (200.0, 0.2969804) (250.0, 0.2955929) (300.0, 0.2961225) (350.0, 0.2983654) (400.0, 0.300025)
    };\addlegendentry{\texttt{DDPG-GM}};
    \addplot[black,mark=triangle,line width=1pt,dashed,mark options=solid] coordinates {
		(100.0, 0.7275254) (150.0, 0.47466207) (200.0, 0.4317108) (250.0, 0.403855) (300.0, 0.37070665) (350.0, 0.3510271) (400.0, 0.33586043)
    };\addlegendentry{\texttt{MAX-GM}};
    \addplot[red,mark=o,line width=1pt,dashed,mark options=solid] coordinates {
		(100.0, 0.75890833) (150.0, 0.5191925) (200.0, 0.47588998) (250.0, 0.4461821) (300.0, 0.40599206) (350.0, 0.38205957) (400.0, 0.3678525)
    };\addlegendentry{\texttt{RND-GM}};
    \addplot[violet,mark=diamond,line width=1pt,dashed,mark options=solid] coordinates {
		(100.0, 0.79841876) (150.0, 0.56909335) (200.0, 0.51989543) (250.0, 0.47950542) (300.0, 0.43509752) (350.0, 0.39599624) (400.0, 0.37089416)
    };\addlegendentry{\texttt{OMA-GM}};
    \end{axis}
    \end{tikzpicture}
  }
  \caption{The average normalized AoI vs. the maximum coverage distance for the $4$-sized network.}
  \label{fig:2b}
\end{figure}
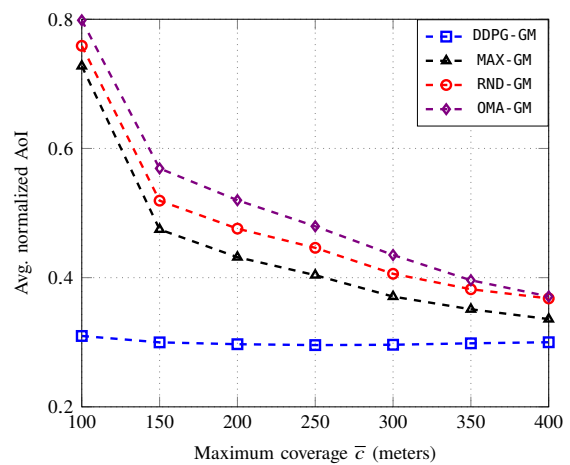

Fig.~\ref{fig:2a} and Fig.~\ref{fig:2b} illustrate the performance of \texttt{DDPG-GM} when the size of the vehicular network and the maximum coverage $\overline{c}$ vary. We have the following conclusions:
\begin{itemize}
    \item \texttt{DDPG-GM} has the lowest AoI;
    \item For \texttt{DDPG-GM}, when the maximum coverage $\overline{c}$ increases up to a certain value, the AoI decreases because a larger $\overline{c}$ implies a larger communication coverage and thus more safety packets will be successfully received;
    \item Beyond a certain value of coverage, the communication coverage of each vehicle increases further, resulting in multiple overlapping communication areas. The improvements of the AoI of the \texttt{DDPG-GM} algorithm is then stopped and it reaches a floor value (and may even start to increase slightly).
\end{itemize}
These conclusions are particularly true for all algorithms in the case of a large network of vehicles. In the case of a smaller vehicular network, we have the following conclusions:
\begin{itemize}
    \item The decrease in AoI is faster, especially for the baselines algorithms. This is due to the sparsity present in the network. In other words, when the number of vehicles is small, choosing a small $\overline{c}$ makes the communication area of each vehicle empty and the AoI will therefore be high. Thus, when $\overline{c}$ increases, the decrease in the AoI will be significant;
    \item In the case of \texttt{DDPG-GM}, due to the efficient learning procedure, the achieved AoI is already minimal even for a small $\overline{c}$, so its decrease when the latter increases are marginal;
    \item We can conclude that increasing the maximum coverage beyond a certain point is unnecessary because the decrease in the AoI is marginal in the \texttt{DDPG-GM} algorithm;
    \item The OMA-based algorithm obtains the highest AoI, which is mainly due to the inefficient use of time slots resources;
    \item The use of the NOMA technique is important for improving the performance of vehicular networks. We further conclude that the \texttt{DDPG-GM} algorithm succeeds in learning better AoI compared to other algorithms thanks to NOMA and the use of the policy gradient-based DRL and DQN approaches.
\end{itemize}

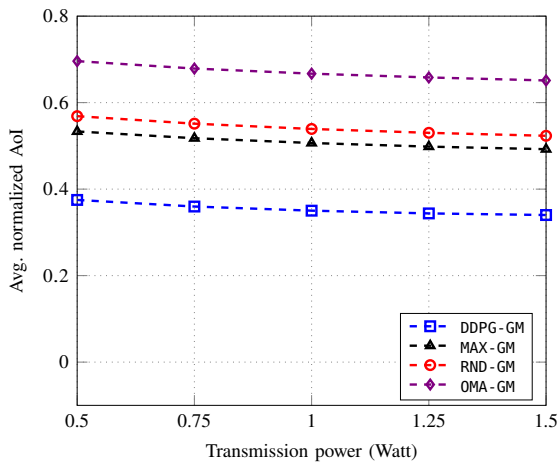
\begin{figure}[h!]
  \centering
  \captionsetup{justification=centering,margin=1cm}
  \resizebox{.42\textwidth}{!}{%
    \begin{tikzpicture}[spy using outlines={circle,magnification=8, connect spies}]
    	\begin{axis}[
		xlabel={Transmission power (Watt)},
        ylabel={Avg. normalized AoI},
        grid=both,
        ymax=0.8,ymin=-0.1,
        xticklabels={0.5, 0.75, 1, 1.25, 1.5},
        xtick={0.5, 0.75, 1, 1.25, 1.5},
        xmin=0.5,xmax=1.5,
        x label style={font=\footnotesize},
        y label style={font=\footnotesize}, 
        ticklabel style={font=\footnotesize},
        legend style={at={(1,1)},font=\scriptsize},]
    \addplot[blue,mark=square,line width=1pt,dashed,mark options=solid] coordinates {
		(0.5, 0.374983) (0.75, 0.359731) (1, 0.350162) (1.25, 0.343843) (1.5, 0.34009) 
    };\label{ddpg10c200p};
    \addplot[black,mark=triangle,line width=1pt,dashed,mark options=solid] coordinates {
		(0.5, 0.533329) (0.75, 0.517693) (1, 0.506615) (1.25, 0.498261) (1.5, 0.49231) 
    };\label{maxmgm10c200p};
    \addplot[red,mark=o,line width=1pt,dashed,mark options=solid] coordinates {
		(0.5, 0.568668) (0.75, 0.551334) (1, 0.539107) (1.25, 0.530157) (1.5, 0.523301) 
    };\label{mgm10c200p};
    \addplot[violet,mark=diamond,line width=1pt,dashed,mark options=solid] coordinates {
		(0.5, 0.69593066) (0.75, 0.6787875) (1, 0.6669472) (1.25, 0.6581924) (1.5, 0.6512414)
    };\label{mgmoma10c200p};
	\end{axis}
    \node [draw,fill=white] at (rel axis cs: 0.83,0.125) {\shortstack[l]{
    \ref{ddpg10c200p} \scriptsize\texttt{DDPG-GM}\\
    \ref{maxmgm10c200p} \scriptsize\texttt{MAX-GM}\\
    \ref{mgm10c200p} \scriptsize\texttt{RND-GM}\\
    \ref{mgmoma10c200p} \scriptsize\texttt{OMA-GM}}};
    \end{tikzpicture}
  }
  \caption{The average normalized AoI vs. the maximum transmission power for the $10$-sized network.}
  \label{fig:3a}
  \end{figure}

In Fig.~\ref{fig:3a} and~\ref{fig:3b}, we illustrate the performance of \texttt{DDPG-GM} when varying the size of the vehicular network and the maximum transmission power $\overline{p}$. We conclude the following:
\begin{itemize}
    \item For all algorithms, the AoI decreases slightly as the maximum transmission power increases since it becomes easier to meet the safety packets requirements with higher transmission powers;
    \item Comparing \texttt{MAX-GM} and \texttt{RND-GM}, we find that the AoI is minimal with \texttt{MAX-GM} because \texttt{RND-GM} uses random power allocation in each NOMA group;
    \item The OMA-based algorithm achieves the highest AoI which is mainly due to the inefficient use of time slots resources. This illustrates the importance of the NOMA technique in vehicular resource allocation.
    \item \texttt{DDPG-GM} manages to achieve better AoI than the other algorithms thanks to NOMA and the use of DRL;
    \item The AoI is higher when the size of the vehicle network is large; which is due to the half-duplex and broadcast nature of V2X communications. Indeed, when there are more vehicles in the network and many of them broadcast their safety packets, the interference becomes high, which affects the performance of the algorithms. Thus, fewer safety packets will be successfully delivered, and consequently, the AoI increases.
\end{itemize}

\begin{figure}[h!]
  \centering
  \captionsetup{justification=centering,margin=1cm}
  \resizebox{.42\textwidth}{!}{%
    \begin{tikzpicture}
    	\begin{axis}[
		xlabel={Transmission power (Watt)},
        ylabel={Avg. normalized AoI},
        grid=both,
        ymax=0.6,ymin=0,
        xticklabels={0.5, 0.75, 1, 1.25, 1.5},
        xtick={0.5, 0.75, 1, 1.25, 1.5},
        xmin=0.5,xmax=1.5,
        x label style={font=\footnotesize},
        y label style={font=\footnotesize}, 
        ticklabel style={font=\footnotesize},
        legend style={at={(1,1)},font=\scriptsize},]
    \addplot[blue,mark=square,line width=1pt,dashed,mark options=solid] coordinates {
		(0.5, 0.3108492) (0.75, 0.3020925) (1.0, 0.2969804) (1.25, 0.2930754) (1.5, 0.29009667)
    };\label{ddpg4c200p};
    \addplot[black,mark=triangle,line width=1pt,dashed,mark options=solid] coordinates {
		(0.5, 0.46234247) (0.75, 0.44464082) (1.0, 0.4317108) (1.25, 0.42251793) (1.5, 0.41640082)
    };\label{maxmgm4c200p};
    \addplot[red,mark=o,line width=1pt,dashed,mark options=solid] coordinates {
		(0.5, 0.5105304) (0.75, 0.4890275) (1.0, 0.47588998) (1.25, 0.46494415) (1.5, 0.45745128)
    };\label{mgm4c200p};
    \addplot[violet,mark=diamond,line width=1pt,dashed,mark options=solid] coordinates {
		(0.5, 0.5495033) (0.75, 0.53228956) (1.0, 0.51989543) (1.25, 0.5109367) (1.5, 0.5041729)
    };\label{mgmoma4c200p};
	\end{axis}
    \node [draw,fill=white] at (rel axis cs: 0.83,0.125) {\shortstack[l]{
    \ref{ddpg4c200p} \scriptsize\texttt{DDPG-GM}\\
    \ref{maxmgm4c200p} \scriptsize\texttt{MAX-GM}\\
    \ref{mgm4c200p} \scriptsize\texttt{RND-GM}\\
    \ref{mgmoma4c200p} \scriptsize\texttt{OMA-GM}}};
    \end{tikzpicture}
  }
  \caption{The average normalized AoI vs. the maximum transmission power for the $4$-sized network.}
  \label{fig:3b}
\end{figure}
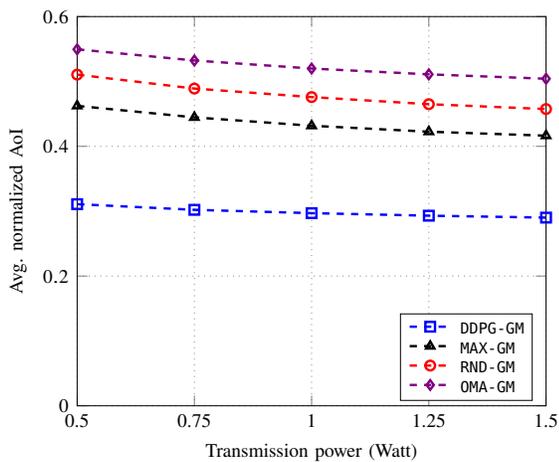

\begin{figure}[h!]
	\centering
	  \captionsetup{justification=centering,margin=1cm}
  \resizebox{.42\textwidth}{!}{%
    \begin{tikzpicture}
    	\begin{axis}[
		xlabel={Maximum coverage (meters)},
        ylabel={Avg. normalized AoI},
        grid=both,
        ymax=1.2,ymin=0.2,
        xticklabels={100,150,200,250,300,350,400},
        xtick={100,150,200,250,300,350,400},
        xmin=100,xmax=400,
        x label style={font=\footnotesize},
        y label style={font=\footnotesize}, 
        ticklabel style={font=\footnotesize},
        legend style={at={(1,1.05)},font=\scriptsize},]
    \addplot[blue,mark=square,line width=1pt,dashed,mark options=solid] coordinates {
		(100.0, 0.450173) (150.0, 0.42320293) (200.0, 0.4103176) (250.0, 0.4071433) (300.0, 0.40600762) (350.0, 0.40783504) (400.0, 0.40938005)
    };\addlegendentry{\texttt{DDPG-GM} with velocity $\mathbf{v}_1$};
    \addplot[black,mark=triangle,line width=1pt,dashed,mark options=solid] coordinates {
		(100.0, 0.6216136) (150.0, 0.5765517) (200.0, 0.558219) (250.0, 0.54410523) (300.0, 0.53494483) (350.0, 0.5306275) (400.0, 0.52761835)            
    };\addlegendentry{\texttt{MAX-GM} with velocity $\mathbf{v}_1$};
    \addplot[red,mark=o,line width=1pt,dashed,mark options=solid] coordinates {
		(100.0, 0.6443441) (150.0, 0.6045301) (200.0, 0.5866476) (250.0, 0.57470775) (300.0, 0.5655407) (350.0, 0.56158185) (400.0, 0.56016654)
    };\addlegendentry{\texttt{RND-GM} with velocity $\mathbf{v}_1$};
    \addplot[violet,mark=diamond,line width=1pt,dashed,mark options=solid] coordinates {
		(100.0, 0.77785856) (150.0, 0.732157) (200.0, 0.7086952) (250.0, 0.6933069) (300.0, 0.6762218) (350.0, 0.66914606) (400.0, 0.65940905)
    };\addlegendentry{\texttt{OMA-GM} with velocity $\mathbf{v}_1$};
    \addplot[blue,mark=square,line width=1pt,dotted,mark options=solid] coordinates {
		(100.0, 0.395043) (150.0, 0.36312526) (200.0, 0.3508745) (250.0, 0.34592277) (300.0, 0.3444209) (350.0, 0.34522477) (400.0, 0.3489156)
    };\addlegendentry{\texttt{DDPG-GM} with velocity $\mathbf{v}_2$};
	\addplot[black,mark=triangle,line width=1pt,dotted,mark options=solid] coordinates {
        (100.0, 0.5913249) (150.0, 0.5487219) (200.0, 0.5327181) (250.0, 0.5218653) (300.0, 0.5152224) (350.0, 0.50459576) (400.0, 0.4825554)
    };\addlegendentry{\texttt{MAX-GM} with velocity $\mathbf{v}_2$};
    \addplot[red,mark=o,line width=1pt,dotted,mark options=solid] coordinates {
		(100.0, 0.61099434) (150.0, 0.57842577) (200.0, 0.5622305) (250.0, 0.55277294) (300.0, 0.5473615) (350.0, 0.5382226) (400.0, 0.5166693)	
    };\addlegendentry{\texttt{RND-GM} with velocity $\mathbf{v}_2$};
    \addplot[violet,mark=diamond,line width=1pt,dotted,mark options=solid] coordinates {
		(100.0, 0.74484515) (150.0, 0.7007709) (200.0, 0.67814195) (250.0, 0.6590293) (300.0, 0.647466) (350.0, 0.63364583) (400.0, 0.6151352)
    };\addlegendentry{\texttt{OMA-GM} with velocity $\mathbf{v}_2$};
    \end{axis}
    \end{tikzpicture}
  }
  \caption{The average normalized AoI vs. the maximum coverage for the $10$-sized network for different speeds of vehicles.}
  \label{fig:4}
\end{figure}
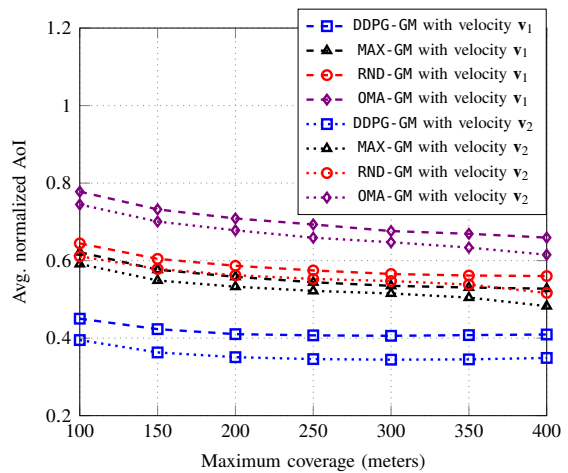%
In Fig.~\ref{fig:4}, we illustrate the impact of vehicle speeds on the AoI by considering a $10$-sized vehicular network. We consider two groups of speeds $\mathbf{v}_1$ and $\mathbf{v}_2$. The $\mathbf{v}_1$ speed is given by $[90, 110, 130, 130, 110, 90]$ and the $\mathbf{v}_2$ speed is given by $[40, 60, 80, 80, 60, 40]$ where the $i$th component of $\mathbf{v}_j$ ($j=1$ or $2$) represents the $i$th lane and the vehicles belonging to this lane move with the corresponding speed. We have the following conclusions:
\begin{itemize}
    \item \texttt{DDPG-GM} always has the best AoI compared to all algorithms and regardless of the speeds;
    \item As the vehicles move faster, minimizing the AoI becomes much more difficult in such a highly dynamic network;
    \item The AoI corresponding to $\mathbf{v}_1$ is higher than the one corresponding to $\mathbf{v}_2$ which is much slower than $\mathbf{v}_1$.
\end{itemize}

\section{Conclusions, Limitations and Future Works}\label{sec:cl}
In this paper, we studied the age of information (AoI) minimization problem in the context of a vehicle-to-everything (V2X) ecosystem by deploying the non-orthogonal multiple access (NOMA) technique. The problem involves coverage optimization, half-duplex transceiver selection, power allocation, and resource blocks (RBs) scheduling. We formulated the problem as a  mixed-integer non-linear problem and studied its NP-hardness. Due to the NP-hardness result of the problem as well as the non-linear optimization formulation, classical optimization techniques are not feasible. To solve this complex problem using reinforcement learning (RL) techniques, we formulated it as a single-agent Markov decision process (MDP). Due to the curse of dimensionality and the mixed nature of the action space that contains discrete actions corresponding to the RBs scheduling and transceiver selection and continuous actions corresponding to coverage selection and power allocation, the application of fingerprint-based deep RL (DRL) approaches, such as the deep-Q-network (DQN) methods, is not feasible. To effectively solve the problem, we proposed a decomposition-based greedy matching and deep deterministic policy gradient (\texttt{DDPG}) algorithm. Specifically, based on previous work, we proposed a stable roommate matching algorithm to solve the RBs scheduling problem. Then the \texttt{DDPG} method is used to learn the continuous coverage and power decisions. The \texttt{DDPG} method is based on recent advances in DQN, including off-policy learning, replay buffer, and batch normalization. We showed that the proposed decomposition-based algorithm outperforms baseline algorithms and successfully provides low AoI in different vehicular network scenarios.

The main limitations of our work that we will address in future works are the following. First, the multi-agent \texttt{DDPG} algorithm could be investigated to solve the continuous resource allocation subproblem and thus implement a fully distributed approach. Second, modeling and solving the entire discrete/continuous resource allocation problem using DRL approach is a challenging task that should be carefully studied in future work where for example the clip and count-based proximal policy optimization could be investigated. Third, a joint method rather than a decomposition-based one would be interesting to investigate in future. Finally, dynamic programming approach can be a perfect candidate solution for the problem that is worth studying.

\bibliographystyle{IEEEtran}
\bibliography{IEEEabrv,M2Mv1}

\begin{thebibliography}{10}
\providecommand{\url}[1]{#1}
\csname url@samestyle\endcsname
\providecommand{\newblock}{\relax}
\providecommand{\bibinfo}[2]{#2}
\providecommand{\BIBentrySTDinterwordspacing}{\spaceskip=0pt\relax}
\providecommand{\BIBentryALTinterwordstretchfactor}{4}
\providecommand{\BIBentryALTinterwordspacing}{\spaceskip=\fontdimen2\font plus
\BIBentryALTinterwordstretchfactor\fontdimen3\font minus
  \fontdimen4\font\relax}
\providecommand{\BIBforeignlanguage}[2]{{%
\expandafter\ifx\csname l@#1\endcsname\relax
\typeout{** WARNING: IEEEtran.bst: No hyphenation pattern has been}%
\typeout{** loaded for the language `#1'. Using the pattern for}%
\typeout{** the default language instead.}%
\else
\language=\csname l@#1\endcsname
\fi
#2}}
\providecommand{\BIBdecl}{\relax}
\BIBdecl

\bibitem{vannithamby20205g}
\BIBentryALTinterwordspacing
R.~Vannithamby \emph{et~al.}, \emph{{5G Verticals: Customizing Applications,
  Technologies and Deployment Techniques}}.\hskip 1em plus 0.5em minus
  0.4em\relax Wiley, 2020. [Online]. Available:
  \url{https://books.google.ca/books?id=mlCatAEACAAJ}
\BIBentrySTDinterwordspacing

\bibitem{9186820}
M.~{Noor-A-Rahim} \emph{et~al.}, ``{A Survey on Resource Allocation in
  Vehicular Networks},'' \emph{{IEEE} Trans. Intell. Transp. Syst.}, pp. 1--21,
  2020, early Access.

\bibitem{7917274}
J.~{Sahoo} \emph{et~al.}, ``{Dynamic Hierarchical Aggregation for Vehicular
  Sensing},'' \emph{{IEEE} Trans. Intell. Transp. Syst.}, vol.~18, no.~9, pp.
  2539--2556, 2017.

\bibitem{3gpp.22.886}
\BIBentryALTinterwordspacing
3GPP, ``{Study on Enhancement of 3GPP Support for 5G V2X Services},'' {3rd
  Generation Partnership Project (3GPP)}, Technical Report (TR) 22.886, 2018,
  version 16.2.0. [Online]. Available:
  \url{https://portal.3gpp.org/desktopmodules/Specifications/SpecificationDetails.aspx?specificationId=3108}
\BIBentrySTDinterwordspacing

\bibitem{8006590}
Y.~{Hsu} \emph{et~al.}, ``{Age of Information: Design and Analysis of Optimal
  Scheduling Algorithms},'' in \emph{Proc. IEEE Int. Symposium on Information
  Theory (ISIT)}, June 2017, pp. 561--565.

\bibitem{10.1145/3331054.3331547}
L.~Baldesi \emph{et~al.}, ``{Keep It Fresh: Reducing the Age of Information in
  V2X Networks},'' in \emph{Proc. ACM MobiHoc Workshop on Technologies, Models,
  and Protocols for Cooperative Connected Cars (TOP-Cars)}, 2019, p. 7–12.

\bibitem{9088326}
S.~A. {Ashraf}, R.~{Blasco}, H.~{Do}, G.~{Fodor}, C.~{Zhang}, and W.~{Sun},
  ``{Supporting Vehicle-to-Everything Services by 5G New Radio Release-16
  Systems},'' \emph{IEEE Commun. Standards Mag.}, vol.~4, no.~1, pp. 26--32,
  2020.

\bibitem{7974737}
B.~{Di} \emph{et~al.}, ``{Non-Orthogonal Multiple Access for High-Reliable and
  Low-Latency V2X Communications in 5G Systems},'' \emph{{IEEE} J. Sel. Areas
  Commun.}, vol.~35, no.~10, pp. 2383--2397, 2017.

\bibitem{6692652}
Y.~{Saito} \emph{et~al.}, ``{Non-Orthogonal Multiple Access (NOMA) for Cellular
  Future Radio Access},'' in \emph{Proc. IEEE Veh. Technol. Conf. (VTC
  Spring)}, 2013, pp. 1--5.

\bibitem{6868214}
Z.~{Ding} \emph{et~al.}, ``{On the Performance of Non-Orthogonal Multiple
  Access in 5G Systems with Randomly Deployed Users},'' \emph{{IEEE} Signal
  Process. Lett.}, vol.~21, no.~12, pp. 1501--1505, 2014.

\bibitem{5984917}
S.~{Kaul} \emph{et~al.}, ``{Minimizing Age of Information in Vehicular
  Networks},'' in \emph{Proc. IEEE Commun. Society Conf. on Sensor, Mesh and Ad
  Hoc Commun. and Networks}, 2011, pp. 350--358.

\bibitem{8937801}
M.~K. {Abdel-Aziz} \emph{et~al.}, ``{Optimized Age of Information Tail for
  Ultra-Reliable Low-Latency Communications in Vehicular Networks},''
  \emph{{IEEE} Trans. Commun.}, vol.~68, no.~3, pp. 1911--1924, 2020.

\bibitem{9205620}
A.~H. Sodhro \emph{et~al.}, ``{Toward ML-Based Energy-Efficient Mechanism for
  6G Enabled Industrial Network in Box Systems},'' \emph{{IEEE} Trans. Ind.
  Informat.}, vol.~17, no.~10, pp. 7185--7192, 2021.

\bibitem{9199859}
------, ``{Towards 5G-Enabled Self Adaptive Green and Reliable Communication in
  Intelligent Transportation System},'' \emph{{IEEE} Trans. Intell. Transp.
  Syst.}, vol.~22, no.~8, pp. 5223--5231, 2021.

\bibitem{9085258}
------, ``{AI-Enabled Reliable Channel Modeling Architecture for Fog Computing
  Vehicular Networks},'' \emph{{IEEE} Wireless Commun.}, vol.~27, no.~2, pp.
  14--21, 2020.

\bibitem{9014535}
------, ``{Link Optimization in Software Defined IoV Driven Autonomous
  Transportation System},'' vol.~22, no.~6, pp. 3511--3520, 2021.

\bibitem{9128850}
Z.~Mlika, O.~Khalifeh, and W.~Ajib, ``Association and scheduling in energy
  harvesting networks: Age of information and fairness trade-off,'' in
  \emph{2020 IEEE 91st Vehicular Technology Conference (VTC2020-Spring)}, 2020,
  pp. 1--5.

\bibitem{8954939}
X.~{Chen} \emph{et~al.}, ``{Age of Information Aware Radio Resource Management
  in Vehicular Networks: A Proactive Deep Reinforcement Learning
  Perspective},'' \emph{{IEEE} Trans. Wireless Commun.}, vol.~19, no.~4, pp.
  2268--2281, 2020.

\bibitem{9195789}
M.~{Samir} \emph{et~al.}, ``{Age of Information Aware Trajectory Planning of
  UAVs in Intelligent Transportation Systems: A Deep Learning Approach},''
  \emph{{IEEE} Trans. Veh. Technol.}, vol.~69, no.~11, pp. 12\,382--12\,395,
  2020.

\bibitem{9214855}
F.~{Peng} \emph{et~al.}, ``{Age of Information Optimized MAC in V2X Sidelink
  via Piggyback-Based Collaboration},'' \emph{{IEEE} Trans. Wireless Commun.},
  vol.~20, no.~1, pp. 607--622, 2021.

\bibitem{8792382}
L.~{Liang}, H.~{Ye}, and G.~Y. {Li}, ``{Spectrum Sharing in Vehicular Networks
  Based on Multi-Agent Reinforcement Learning},'' \emph{{IEEE} J. Sel. Areas
  Commun.}, vol.~37, no.~10, pp. 2282--2292, 2019.

\bibitem{8943940}
L.~{Liang}, H.~{Ye}, G.~{Yu}, and G.~Y. {Li}, ``{Deep-Learning-Based Wireless
  Resource Allocation With Application to Vehicular Networks},'' \emph{Proc.
  {IEEE}}, vol. 108, no.~2, pp. 341--356, 2020.

\bibitem{mlika2021network}
Z.~Mlika and S.~Cherkaoui, ``Network slicing for vehicular communications: a
  multi-agent deep reinforcement learning approach,'' \emph{Annals of
  Telecommunications}, vol.~76, no.~9, pp. 665--683, 2021.

\bibitem{9318243}
------, ``Network slicing with mec and deep reinforcement learning for the
  internet of vehicles,'' \emph{IEEE Network}, vol.~35, no.~3, pp. 132--138,
  2021.

\bibitem{9524882}
A.~Abouaomar, Z.~Mlika, A.~Filali, S.~Cherkaoui, and A.~Kobbane, ``A deep
  reinforcement learning approach for service migration in mec-enabled
  vehicular networks,'' in \emph{2021 IEEE 46th Conference on Local Computer
  Networks (LCN)}, 2021, pp. 273--280.

\bibitem{7852321}
I.~{Kadota}, E.~{Uysal-Biyikoglu}, R.~{Singh}, and E.~{Modiano}, ``{Minimizing
  the Age of Information in Broadcast Wireless Networks},'' in \emph{Proc. ACM
  Allerton}, 2016, pp. 844--851.

\bibitem{10.1145/3209582.3209602}
N.~Lu, B.~Ji, and B.~Li, ``{Age-Based Scheduling: Improving Data Freshness for
  Wireless Real-Time Traffic},'' in \emph{Proc. ACM MOBIHOC}, 2018, p.
  191–200.

\bibitem{8632657}
M.~{Zeng}, A.~{Yadav}, O.~A. {Dobre}, and H.~V. {Poor}, ``{Energy-Efficient
  Joint User-RB Association and Power Allocation for Uplink Hybrid NOMA-OMA},''
  \emph{{IEEE} Internet Things J.}, vol.~6, no.~3, pp. 5119--5131, 2019.

\bibitem{Garey:1979}
M.~R. Garey and D.~S. Johnson, \emph{Computers and Intractability: A Guide to
  the Theory of NP-Completeness}.\hskip 1em plus 0.5em minus 0.4em\relax New
  York, NY, USA: W. H. Freeman \& Co., 1979.

\bibitem{8904331}
Z.~{Mlika} \emph{et~al.}, ``{Packet Scheduling Algorithms to Minimize the Age
  of Information in Energy Harvesting Networks},'' in \emph{Proc. IEEE Annual
  Int. Symposium on Personal, Indoor and Mobile Radio Commun. (PIMRC)}, 2019,
  pp. 1--6.

\bibitem{lillicrap2019continuous}
T.~P. Lillicrap \emph{et~al.}, ``Continuous control with deep reinforcement
  learning,'' 2019.

\bibitem{10.5555/68392}
D.~Gusfield and R.~W. Irving, \emph{{The Stable Marriage Problem: Structure and
  Algorithms}}.\hskip 1em plus 0.5em minus 0.4em\relax Cambridge, MA, USA: MIT
  Press, 1989.

\bibitem{3gpp.37.885}
\BIBentryALTinterwordspacing
3GPP, ``{Study on Evaluation Methodology of New Vehicle-to-Everything (V2X) Use
  Cases for LTE and NR},'' {3rd Generation Partnership Project (3GPP)},
  Technical Report (TR) 37.885, 06 2019, version 15.3.0. [Online]. Available:
  \url{https://portal.3gpp.org/desktopmodules/Specifications/SpecificationDetails.aspx?specificationId=3209}
\BIBentrySTDinterwordspacing

\bibitem{bezanson2017julia}
\BIBentryALTinterwordspacing
J.~Bezanson, A.~Edelman, S.~Karpinski, and V.~B. Shah, ``Julia: A fresh
  approach to numerical computing,'' \emph{SIAM review}, vol.~59, no.~1, pp.
  65--98, 2017. [Online]. Available: \url{https://doi.org/10.1137/141000671}
\BIBentrySTDinterwordspacing

\bibitem{innes:2018}
M.~Innes, ``Flux: Elegant machine learning with julia,'' \emph{Journal of Open
  Source Software}, 2018.

\bibitem{winner}
\BIBentryALTinterwordspacing
P.~Kyosti \emph{et~al.}, ``\BIBforeignlanguage{English}{{WINNER II Channel
  Models: document IST-4-027756 WINNER II D1.1.2 V1.2}},'' Tech. Rep., Sep.
  2007. [Online]. Available:
  \url{https://pdfs.semanticscholar.org/dd24/b09f0f95367ce73a2a1445445a802556ac5e.pdf}
\BIBentrySTDinterwordspacing

\bibitem{adam}
D.~P. {Kingma} and J.~{Ba}, ``{Adam: A Method for Stochastic Optimization},''
  \emph{arXiv e-prints}, p. arXiv:1412.6980, Dec. 2014.

\end{thebibliography}

\end{document}